\documentclass[reqno]{article}

\usepackage{cite}
\usepackage{amssymb,amsfonts}
\usepackage[capposition=top]{floatrow}
\usepackage[all,arc]{xy}
\usepackage{enumerate}
\usepackage{mathrsfs}
\usepackage{parskip}
\usepackage{amsfonts}
\usepackage{amsmath}
\usepackage{tikz}
\usepackage[normalem]{ulem}
\usetikzlibrary{calc}
\usepackage{cancel}
\usepackage{mathabx}
\usepackage{setspace}
\usepackage{graphicx}
\DeclareGraphicsExtensions{.png}
\DeclareGraphicsExtensions{.pdf}
\usepackage{float}
\usepackage[margin=1in]{geometry}
\numberwithin{equation}{section}
\usepackage{dashrule}
\usepackage{mathtools}
\usepackage{bm}
\usepackage{tikz}
\usetikzlibrary{shapes,arrows}
\usepackage{subcaption}
\usepackage[authoryear]{natbib} 
\usepackage{cancel}
\usepackage{multirow}

\usepackage{authblk}

\usepackage{pgfplots}
\usetikzlibrary{arrows.meta}
\usetikzlibrary{patterns}
\tikzset{>={Latex[width=1.5mm,length=1.5mm]}}

\begin{document}

\title{Impedance response of ionic liquids in long slit pores}

\author[1]{Ruben J. Tomlin}
\affil[1]{Department of Mechanical Engineering, Imperial College London, London SW7 2AZ, UK}

\author[2]{Tribeni Roy}
\affil[2]{Department of Mechanical Engineering, BITS Pilani, India}

\author[3]{Toby L. Kirk}
\affil[3]{Mathematical Institute, University of Oxford, Oxford OX2 6GG, UK}

\author[1]{Monica Marinescu}

\author[4]{Dirk Gillespie}
\affil[4]{Department of Physiology and Biophysics, Rush University Medical Center, Chicago, IL USA}

\maketitle





\begin{abstract}

{\color{red}

}

We study the dynamics of ionic liquids in a thin slit pore geometry. Beginning with the classical and dynamic density functional theories for systems of charged hard spheres, an asymptotic procedure leads to a simplified model which incorporates both the accurate resolution of the ion layering (perpendicular to the slit pore wall) and the ion transport in the pore length. 
This reduced-order model enables qualitative comparisons between different ionic liquids and electrode pore sizes at low numerical expense. We derive semi-analytical expressions for the impedance response of the reduced-order model involving numerically computable sensitivities, and obtain effective finite-space Warburg elements valid in the high and low frequency limits. Additionally, we perform time-dependent numerical simulations to recover the impedance response as a cross-validation step. We investigate the dependence of the impedance response on system parameters and the choice of density functional theory used. The inclusion of electrostatic effects beyond mean-field qualitatively changes the dependence of the characteristic response time on the pore width. We observe peaks in the response time as a function of pore width, with height and location depending on the potential difference imposed. We discuss how the calculated dynamic properties can be used together with equilibrium results to optimise ionic liquid supercapacitors for a given application.

\end{abstract}

\section{Introduction}

Supercapacitors are typically comprised of porous carbon-derived electrodes with an aqueous, organic, or ionic liquid electrolyte, and store charge in electrostatic double layers at all surfaces in the pore network. Models have been constructed at a varying levels of complexity to describe this system and its components, i.e.~individual pores, pore networks or electrodes. For short individual pores, it can be viable to perform molecular dynamics simulations to resolve the complicated many-body interactions in full, although this is often very computationally expensive. At the cell level, physics-based models have been developed with an averaging approach under the assumptions of porous electrode and dilute solution theories. These can incorporate electrolyte transport at multiple pore scales \citep{verbrugge2005microstructural,kroupa2016modelling}.

For concentrated electrolytes, such as ionic liquids, and systems where the pore size and electrostatic double layer widths are comparable, Poisson--Boltzmann (PB) theory for dilute solutions becomes invalid. More complicated models have been developed, incorporating ion size (steric) effects and electrostatics beyond the mean-field description. Concerned with the dynamics of concentrated electrolytes, \cite{kilic2007steric1} considered two models for an electrostatic double layer, the modified PB model of \cite{borukhov1997steric} which includes steric effects in a mean-field manner, and a composite model incorporating an inner layer of ions at the electrode surface with high packing fraction and a dilute PB outer layer. \cite{kilic2007steric1} showed that steric effects suppress the unbounded growth of the zero-frequency differential capacitance ($\mathrm{d}\mathcal{Q}/\mathrm{d}\mathcal{V}$) with voltage found in PB theory, exhibiting bell- or camel-shaped curves where either one or two maxima are present, respectively -- this finding was corroborated by \cite{doi:10.1021/jp067857o} with theoretical results for a lattice-gas model, also showing that different ion sizes would result in the asymmetry of $\mathrm{d}\mathcal{Q}/\mathrm{d}\mathcal{V}$ about zero potential (relative to the potential in the bulk). At the level of equivalent circuit models, \cite{kilic2007steric1} then predict that the dynamical response time of the system (which scales linearly with the capacitance) is greatly reduced from the PB result at large voltages.

In a companion paper, \cite{kilic2007steric2} derive a modified Poisson--Nernst--Planck (PNP) system based on the free energy functional of \cite{borukhov1997steric}, and perform numerical simulations of a parallel-plate capacitor. When charging, the double layer governed by the modified PNP system initially proceeds as it would with the unmodified PNP equations, until the packing limit is reached and the double layer extends laterally with constant ion concentrations (while the unmodified PNP solution grows exponentially at the electrode surface). This is however a qualitatively incorrect description of how steric effects affect double layer structure \citep{gillespie2015review}. A phenomenological model with electrostatic correlations, improving on the usual mean-field electrostatics, was proposed by \cite{bazant2011double}. At low voltages, overscreening/charge inversion was observed, where the magnitude of charge on the electrode surface is less than that in the first layer of counter-ions.

\cite{kondrat2013charging} studied pore charging using an improved mean-field model with electrostatic correlations and the steric free energy of \cite{borukhov1997steric}, solving a spatially 2D continuity equation in long slit pores of width and length on the nanometer scale. In most of their simulations, they found that wider pores charged faster, and observed a front-like propagation of charge into empty pores, with initially filled pores charging diffusively. Additionally, \cite{kondrat2013charging} outlined an approach for optimising a porous electrode with a simplified schematic in which the pore widths and lengths are tuned to attain the best combination of energy and power density.

Another approach to model the behaviour of confined and concentrated electrolytes is classical density functional theory (DFT). DFT computes the equilibrium behaviour of inhomogeneous fluids in the presence of external potentials by minimising a free energy functional \citep{evans1979nature} (the reviews of \cite{evans1992density} and \cite{lutsko2010recent} provide an overview of the major advances in classical DFT since its inception). 
In DFT, ionic liquids are typically modelled as mixtures of charged hard spheres; for these interactions, good approximate free energy functionals exist.
The hard sphere (steric) component is incorporated using the fundamental measure theory (FMT) derived by \cite{rosenfeld1989free}, or the improved White Bear versions constructed by \cite{yu2002structures,roth2002fundamental} and \cite{hansen2006density}. Theories for electrostatic correlations, based on the mean spherical approximation (MSA) for bulk systems \citep{blum1975mean}, have been derived by a number of authors \citep{kierlik1991density,rosenfeld1993free,gillespie2002coupling,gillespie2003density,roth2016shells}. Both the hard sphere and electrostatic theories used in the present work are discussed below.

Classical DFT has already been utilised extensively to analyse ionic liquid choice and pore size effects with applications in supercapacitor design. However, DFT is limited to the computation of equilibrium quantities such as the integral capacitance or $\mathrm{d}\mathcal{Q}/\mathrm{d}\mathcal{V}$. \cite{jiang2011oscillation} observed that the integral capacitance can have a strongly oscillatory dependence on the pore width, indicating that the optimisation of carbonaceous electrode porosities could lead to a significant increase in the energy density of an ionic liquid supercapacitor. The influence of bulk ion densities and ion size asymmetry on $\mathrm{d}\mathcal{Q}/\mathrm{d}\mathcal{V}$ was investigated by \cite{jiang2011density}, observing both camel- and bell-shape plots for the differential capacitance as a function of the pore surface potential for low and high bulk densities, respectively, as found in previous studies incorporating ion size effects \citep{kilic2007steric1,doi:10.1021/jp067857o}. Further work has focused on the influence of an added solvent \citep{jiang2012solvent} or other impurities \citep{liu2017impurity}, and the possibility of mixing ionic liquids to maximise capacitance \citep{lian2016enhancing,neal2017ion}.

Dynamic density functional theory (DDFT) extends DFT to non-equilibrium systems. \cite{evans1979nature} was the first to write down a deterministic dynamical equation which governs the evolution of the particle densities, with later derivations due to \cite{dieterich1990nonlinear} and \cite{marconi1999dynamic}. The DDFT equation has the form of a conservation law, and specifies that particle density distributions evolve according to gradients in the local chemical potentials (these are spatially varying out of equilibrium), relaxing to classical DFT equilibria. The validity of this model far from equilibrium is being investigated \citep{schmidt2013power}. \cite{te2020classical} provide an overview of the original DDFT model of \cite{evans1979nature} and its extensions, with a comprehensive survey of the many applications. The DDFT system reduces to the usual diffusion equation in the case of a non-interacting ideal gas, and classical PNP can be obtained for dilute charged particle systems with a mean-field assumption for the electrostatics.

There have been a number of DDFT studies for ionic liquids; the majority of these consider a nanoscale parallel plate capacitor set-up, such as \cite{jiang2014time,lian2016time} and \cite{qingsurface}, who studied charging dynamics. \cite{babel2018impedance} applied a sinusoidal voltage on the walls of a nanoscale parallel-plate capacitor to calculate the impedance response, performing numerical simulations both with and without mean-field electrostatics and hard-sphere effects, finding that the latter had the most significant influence on the result. Also of relevance to the present study is the work of \cite{qing2019dynamic} on ion dynamics in like-charged nano-slits. The authors fixed the density and chemical potential equal to bulk values at permeable pore walls, and allowed the slit to fill from empty, with a focus on determining the adsorption capacity and rate as a function of the pore width. All of the above works are spatially 1D, and capture the fast dynamics of ionic liquids on the nanoscale.

In the present work, we are interested in using DFT and DDFT to predict the influence of the underlying ion equilibria on the impedance response. The study of impedance response of porous electrodes was pioneered by \cite{de1963porous,de1964porous,de1967electrochemical} who modelled the system as a set of infinite and semi-infinite cylindrical pores connected in parallel. He proposed a transmission line (electrical circuit) model for the ion transport in the pores, for which the circuit parameters could be expressed in terms of physical quantities. Under a number of simplifying assumptions, including electroneutrality, infinite electrical conductivity in the electrode (valid for carbon-based supercapacitor electrodes), and dominance of ion transport by migration over diffusion, the impedance response of a single semi-infinite pore homogeneously filled with electrolyte can be reduced to the finite-space Warburg (FSW) element defined by
\begin{equation}\label{deLeviemodel} Z(\omega) = \mathcal{Z}  \frac{\coth ( \sqrt{ \textrm{i} \omega \tau} ) }{\sqrt{ \textrm{i} \omega \tau}}.\end{equation}
Here, $\textrm{i} = \sqrt{-1}$ and $\omega$ is the angular frequency of the alternating voltage/current input, $\mathcal{Z}$ is an impedance and $\tau$ is a time constant. The high and low frequency limits of \eqref{deLeviemodel} are
\begin{equation}\label{deLeviemodellimits} Z(\omega) \xrightarrow{{\omega \rightarrow 0}} \mathcal{Z} \left( \frac{1}{3} - \frac{ \textrm{i}}{ \omega \tau} \right), \qquad  Z(\omega) \xrightarrow{ {\omega \rightarrow \infty}}  \frac{\mathcal{Z}(1-\textrm{i})}{\sqrt{2 \omega \tau}},\end{equation}
respectively \citep{song2012effects}. The high frequency limit pertains to diffusive behaviour, and at low frequencies the system is capacitive.

Analytical expressions similar to \eqref{deLeviemodel} can be derived if some of the assumptions used by de Levie are relaxed, as discussed in the review of \cite{huang2020review}. However, there has been only limited success in incorporating more complex physical models into the analytical derivations of impedance response expressions. For example, in the planar electrode geometry, \cite{li2021impedance} derived the analytical expression for the impedance response using the PNP model without the electroneutrality assumption, whereas \cite{wang2012intrinsic} resorted to a numerical investigation of the same set-up with the PNP model and its modifications due to \cite{LIM2007159} and \cite{kilic2007steric2} (see also \cite{ROLING2012526} and \cite{WANG2012529}). Experimental works often attempt to fit data to \eqref{deLeviemodel}, replacing the exponent $1/2$ with an unknown (constant phase element) exponent if needed. The influence of non-trivial pore geometry was first investigated by \cite{keiser1976abschatzung}, who utilised the transmission line model of de Levie and varied the circuit parameters to account for spatial dependence of the cylindrical pore radius. This problem was recently revisited by \cite{cooper2017simulated} who performed numerical simulations of a diffusion equation in frequency space, both in open and closed pores of varying cross section and in more complicated and tortuous geometries including fractal domains. In the present work, our interest is solely in the influence of improvements to the physical model employed, keeping the geometry as simple as possible.

We study ion dynamics in idealised, spatially 2D, rectangular slit pores, for which the pore length is significantly larger than the nanoscale width. The pore is connected to an infinite particle bath, and the system is driven from equilibrium by variations of the potential on the slit walls. Due to the multi-scale nature of this problem, it would likely be intractable to solve a full 2D DDFT, let alone perform molecular dynamics simulations. We proceed with an asymptotic reduction of the DDFT system, resulting in a limit model on the timescale of supercapacitor applications for which the ion densities appear to equilibrate instantaneously in the nanoscale pore width. This model is the simplest way of incorporating both DFT, for the accurate resolution of the (out-of-plane) ion profiles, and time-dependent dynamics for the ion transport in the pore length dimension. During the preparation of this manuscript, we became aware of the work of \cite{aslyamov2020relation} who derive the same limit model to study the charging of slit pores.

We perform a frequency response analysis of the limit model, finding that it is amenable to a semi-analytical approach. We recover the FSW element \eqref{deLeviemodel} in the case where only one ion species is present in the pore, and construct effective FSW elements, valid in the low and high frequency limits, for the general case. Additionally, we carry out fully numerical simulations in which the time-dependent system is solved in response to a sinusoidal voltage input to mimic a potentiostatic electrochemical impedance spectroscopy (PEIS) experiment at the pore scale (akin to the study of \cite{wang2012intrinsic}). Exploration of the system response beyond the linear PEIS regime through full numerical simulation is left to a later study, although relevant results with discontinuous voltage inputs are presented in \cite{aslyamov2020relation}. For two parameter sets inspired by real ionic liquids, we investigate the influence of system parameters and the choice of DFT. We find that the inclusion of electrostatic correlations in the DFT modify the dynamic characteristics of the pore system for widths close to those of the ions. Without electrostatic correlations, the characteristic time of the linear response appears to be overestimated for pores that can fit a single ion. We observe peaks in the response time as a function of pore width, which shift in magnitude and location depending on the base voltage. For the optimisation of supercapacitor geometries, such dynamic results should be taken together with equilibrium results to ensure that an appropriate balance of energy and power density is achieved for the desired application.

In Section \ref{problemform}, we outline the governing equations for the simplified pore system and discuss the asymptotic model reduction. We provide the semi-analytical theory and outline the fully numerical approach for the impedance response analysis in Section \ref{ImpRes}, and present results for two sets of parameters in Section \ref{numresults}. A discussion and our concluding remarks are contained in Section \ref{Conclu}.

\section{Problem formulation and model reduction \label{problemform}}

\begin{figure}
\begin{tikzpicture}[scale=2.75]

\draw[thick]
(0,0) -- (2,0)
(0,0.5) -- (2,0.5)
(1,0) node[below] {$\psi = \mathcal{V}(t)$}
(1,0.5) node[above] {$\psi =\mathcal{V}(t)$};

 \draw[thick]
(0-0.5,0)  -- (2+0.5,0) coordinate (A4)
(0-0.5,0.5)  -- (2+0.5,0.5) coordinate (A3);

 \draw[thick]
(2+0.5,0) -- (2+0.5,0.5) ;

 \draw[thick]
(0-0.5,0) coordinate (A5) -- (0-0.5,-0.25) coordinate (A6)
(0-0.5,0.5) coordinate (A2) -- (0-0.5,0.5+0.25) coordinate (A1);

 \draw[<->]
  (-0.25,0) -- (-0.25,0.5);
  \draw[thick]
  (-0.25,0.25) node[left] {${L}$};

 \draw[<->]
 (0-0.5,-0.35) --  (2+0.5,-0.35);
  \draw[thick]
  (1,-0.35) node[below] {${H}$};

\draw[thick]
(1,0.25) node {$\rho_i(x,y,t), \quad \mu_i(x,y,t), \quad \psi(x,y,t)$}
(-1.25,0.35) node {$\rho_i^{\textrm{b}}, \quad \mu_i^{\textrm{b}}$,}
(-1.25,0.15) node {$\psi = 0$};

         \def\mypath{ (A1) -- (A2) -- (A3) -- (A4) -- (A5) -- (A6) --  (2+0.75,-0.25) -- (2+0.75,0.75)}
         
           \fill[gray,opacity=0.1] \mypath;
            \fill[pattern=north east lines,opacity=0.1] \mypath;

  \draw[->] (2+0.75+0.25+0.25,0.125) -- (2+1+0.25+0.25,0.125) node[right] {$x$};
  \draw[->] (2+0.75+0.25+0.25,0.125) -- (2+0.75+0.25+0.25,0.25+0.125) node[above] {$y$};

\end{tikzpicture}
\caption{A single slit pore connected to an infinite particle bath.}\label{schematic_main}
\end{figure}
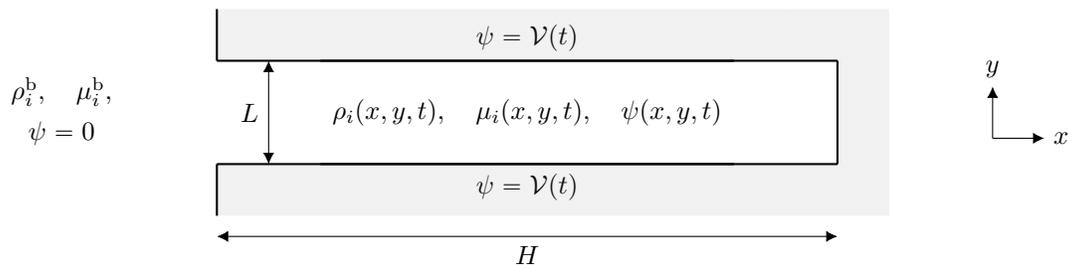

We consider the dynamics of ionic liquids in slit pores of length ${H}$ and width ${L}$ (with an infinite third ``depth" dimension) as shown in the schematic in Figure \ref{schematic_main}. Our interest is in developing models for thin pores with ${H} \gg {L}$, where ${L}$ is on the order of nanometers (comparable to typical ion sizes). The coordinates in the pore length and width are $x$ and $y$, respectively, and in the following discussion we use $\bm{r}$ to denote a general coordinate vector. This individual slit pore represents one of many that comprise a typical carbonaceous electrode in an electrostatic double layer capacitor, specifically of nano-wall type. The filling ionic liquid is modelled as a mixture of hard spheres with point charges located at the particle centres, referred to as the primitive model; specifically, the ion--ion pair potential between particles of species $i,j$ with separation distance $r = |\bm{r}|$ is
\begin{equation}\label{interactionpot1}u_{i,j}(r) = \begin{cases}
            \infty & \text{if $r < R_i + R_j$}, \\
            \displaystyle{ \frac{ z_i z_j {\textrm{e}}^2}{4 \pi \epsilon_0 \epsilon_{\textrm{r}} r } }& \text{if $r \geq R_i + R_j$},
        \end{cases}\end{equation}
where $R_i = \sigma_i/2$ and $z_i$ denote the radius and valence of the $i^{\textrm{th}}$ particle species, respectively. Additionally, ${\textrm{e}}$ is the elementary charge, $\epsilon_0$ denotes the permittivity of free space, and $\epsilon_{\textrm{r}}$ is the dielectric constant of the medium. The electrode material is assumed to have the same dielectric constant to avoid image charges.

The time evolution of the ion density distributions in the pore, denoted $\rho_i(\bm{r},t)$, are governed by the deterministic DDFT system first written down by \cite{evans1979nature},
\begin{equation}\label{DDFT3D}
\frac{\partial \rho_i}{\partial t} = \beta D_i \bm{\nabla} \bm{\cdot} \left( \rho_i \bm{\nabla} \mu_i \right).
\end{equation}
Here, $\mu_i$ are the local chemical potentials, which are spatially varying out of equilibrium. Gradients in the local chemical potentials drive the density evolution, and without external forcing, solutions of \eqref{DDFT3D} relax to equilibrium DFT solutions with constant $\mu_i$. The product $\beta D_i$ is the mobility of the $i^{\textrm{th}}$ species, where $D_i$ denotes a single-particle self-diffusion coefficient and $\beta = 1/k_{\textrm{B}}T$ is the thermodynamic inverse temperature (both $\beta$ and the $D_i$ are assumed constant in this work). At the pore cap we impose a no-flux boundary condition, $\partial_x \mu_i = 0$ at $x=H$, as well as on the long sides of the slit pore, $\partial_y \mu_i = 0$. At $x=0$, the pore is open and connected to an infinite particle bath, with bulk densities and chemical potentials in the far field denoted by $\rho_i^{\textrm{b}}$ and $\mu_i^{\textrm{b}}$, respectively. When the system is at rest, the chemical potentials in the pore equal those in the bulk, $\mu_i = \mu_i^{\textrm{b}}$. The relationship between the densities and local chemical potentials, required to close the system, is based on that used in classical DFT for particle systems in equilibrium, as described next.

The pair interaction potential \eqref{interactionpot1} uniquely determines an intrinsic Helmholtz free energy, $F[ \{\rho_i\} ]$, a functional of the particle densities \citep{evans1979nature}. This free energy is independent of the external potentials, denoted ${{U}}_i$, which describe the influence of the pore walls on species $i$. The grand potential energy density functional for the system is given by
\begin{equation}
\label{grandpotential}
 \Omega[\{\rho_i\}] = F[\{\rho_i\}] + \sum_{i} \int  ({{U}}_i - \mu_i)\rho_i \; \mathrm{d}\bm{r}.
\end{equation}
Equilibrium DFT solutions are obtained as minimisers of $\Omega$, satisfying the Euler--Lagrange equations,
\begin{equation}\label{EL0}
0 = \frac{\delta \Omega}{\delta \rho_i} = \frac{\delta F}{\delta \rho_i} + {{U}}_i - \mu_i,
\end{equation}
which is also used to relate the constant bulk quantities $\rho_i^{\textrm{b}}$ and $\mu_i^{\textrm{b}}$ (for which $U_i = 0$). The DDFT assumption is that \eqref{EL0} remains valid close to equilibrium, and so yields the desired expression for the local chemical potentials to be used in \eqref{DDFT3D}. While there exists a unique Helmholtz free energy corresponding to the particle interactions \eqref{interactionpot1}, it cannot be written down exactly (analytical expressions for $F$ only exist for special cases of $u_{i,j}$). For an exact Helmholtz free energy, it can be shown that there is a unique correspondence between the external potential and the equilibrium density \citep{evans1979nature}, as the minimisation problem is convex, yielding a unique global minimum (unique equilibrium). The DFT approach relies critically on the ability to construct accurate approximations of $F$. However, the convexity property of the minimisation problem can be lost during this approximation process, and sometimes more than one equilibrium solution exists for a given external potential with an approximate free energy \citep{te2020classical}.

Typically, $F$ is decomposed into ideal and excess components, and for the case of pair potential \eqref{interactionpot1}, the excess free energy is split up into contributions due to hard-sphere and electrostatic interactions, with the latter broken down further into mean-field direct Coulomb interactions and electrostatic correlations (sometimes referred to as the electrostatic residual or screening contribution),
\begin{equation}\label{freeenergysplit}F = F^{\textrm{ID}} + F^{\textrm{EX}} = F^{\textrm{ID}} + F^{\textrm{HS}} + F^{\textrm{CO}} + F^{\textrm{EC}}.\end{equation}
The ideal component in \eqref{freeenergysplit} corresponds to the intrinsic Helmholtz free energy for a system of noninteracting particles, i.e.~$u_{i,j} \equiv 0$, and is given exactly by
\begin{equation}\label{idealfreeenergy}F^{\textrm{ID}}[\{\rho_i\}]  = \beta^{-1} \sum_{i} \int ( \log ( \Lambda_i^3 \rho_i(\bm{r}) ) - 1)\, \rho_i(\bm{r})\; \mathrm{d}\bm{r}, \end{equation}
where $\Lambda_i$ is an effective (de Broglie) thermal wavelength \citep{tarazona2008density}. If $F^{\textrm{EX}} = U_i = 0$, the local chemical potential is given by $\mu_i = \beta^{-1} \log \rho_i$ and returns the standard diffusion equation when substituted into \eqref{DDFT3D}. Substituting (\ref{freeenergysplit},\ref{idealfreeenergy}) into the Euler--Lagrange equations \eqref{EL0} gives
\begin{equation}\label{expform1} \rho_i(\bm{r})  = \Lambda_i^{-3} \exp\left[ - \beta \left( \frac{\delta F^{\textrm{EX}}}{\delta \rho_i} + {{U}}_i(\bm{r})  - \mu_i \right) \right].
\end{equation}
This implicit expression for the densities is commonly used as the basis of Picard iteration numerical schemes for equilibrium DFT solutions. The $\Lambda_i$ can be absorbed into the chemical potentials or, at equilibrium, can be cancelled by relating the bulk densities and chemical potentials \citep{tang2003density}.

The direct Coulomb contribution to the excess free energy is
\begin{equation}F^{\textrm{CO}} = \frac{1}{2} \frac{{\textrm{e}}^2}{4\pi \epsilon_0\epsilon_{\textrm{r}}} \sum_{i,j} \int \int  \frac{z_i  z_j \rho_i(\bm{r}) \rho_j(\bm{r'}) }{ |\bm{r} - \bm{r'}|} \; \mathrm{d}\bm{r'} \mathrm{d}\bm{r}.
\end{equation}
This is the leading order term of the expansion of the excess free energy functional corresponding to the pair interaction $u_{i,j}(r) = {\textrm{e}}^2 z_i z_j/4 \pi \epsilon_0 \epsilon_{\textrm{r}} r$, and comprises the electrostatics employed in the PB theory valid for weak electrolytes. This mean-field expression treats the particles as fully uncorrelated, only accounting for the potential created on a given particle by the mean particle distribution \citep{evans1992density,tarazona2008density}. The external potential we consider can be decomposed into the sum of hard-wall and electrostatic components, $U_i = U_i^{\textrm{HW}} + U_i^{\textrm{ES}}$, where $U_i ^{\textrm{ES}}$ can be identified as the the electrostatic potential energy of an ion of charge ${\textrm{e}}z_i$ in the presence of the electrostatic potential generated by all non-ionic charges in the system (e.g.~charged walls). The mean electrostatic potential $\psi(\bm{r})$, incorporating the electrostatic components due to ionic and non-ionic charges \citep{wang2011weighted,wu2011classical,medasani2014ionic}, is defined by
\begin{equation}\label{electrostaticincorporation}{\textrm{e}} z_i \psi(\bm{r}) = \frac{\delta F^{\textrm{CO}}}{\delta \rho_i}  + U_i ^{\textrm{ES}}(\bm{r}) = {\textrm{e}} z_i  \sum_{j} \int  \frac{ {\textrm{e}} z_j \rho_j(\bm{r'}) }{4\pi \epsilon_0\epsilon_{\textrm{r}} |\bm{r} - \bm{r'}|} \; \mathrm{d}\bm{r'} + U_i ^{\textrm{ES}}(\bm{r}).
\end{equation}
The electrostatic potential satisfies the Poisson equation in the ionic liquid phase (the non-ionic component is harmonic in this region),
\begin{equation}\label{Poisson0}\epsilon_0\epsilon_{\textrm{r}} \nabla^2\psi = - {\textrm{e}} \sum_i z_i \rho_i.\end{equation}
As indicated in Figure \ref{schematic_main}, we impose a voltage (measured relative to zero potential in the bulk) at the pore walls, 
   \begin{equation}\label{PoissonBC1}\psi|_{y=0} = \psi|_{y={L}} = \mathcal{V}(t),\end{equation}
   which we use to drive the system from equilibrium.
Without contributions to the free energy from hard-sphere effects and electrostatic correlations, the local chemical potential becomes $\mu_i = \beta^{-1} \log \rho_i + {\textrm{e}}z_i \psi$. This recovers Nernst--Planck when substituted into \eqref{DDFT3D}.

For the hard sphere component, $F^{\textrm{HS}}$, we use the White Bear II functional developed by \cite{hansen2006density}, improving on the original FMT functional of \cite{rosenfeld1989free}. We also performed computations with FMT and White Bear I \citep{roth2002fundamental,yu2002structures}, finding that the influence of this choice was negligible.
Further discussion of the various hard-sphere functionals can be found in the review by \cite{roth2010fundamental}.

The earliest approximation of the electrostatic correlation contribution was derived independently by \cite{kierlik1991density} and \cite{rosenfeld1993free}. They used a functional Taylor expansion about constant bulk densities with coefficients (direct correlation functions) given by the MSA. This bulk fluid density (BFD) approach was refined by \cite{gillespie2002coupling,gillespie2003density} who used self-consistent non-homogeneous reference densities in the Taylor expansion, termed reference fluid density (RFD) theory, with the resulting equilibrium profiles often in close agreement to molecular simulations \citep{voukadinova2018assessing}. More recently, functionalised mean spherical approximation (fMSA) was developed by \cite{roth2016shells} which significantly reduces the computational complexity with only some loss of accuracy (compared to RFD) \citep{voukadinova2018assessing}. In the present work, we show results for both BFD and fMSA theories.
In their study of pore charging, \cite{aslyamov2020relation} use FMT and the weighted correlation approach (for $F^{\textrm{EC}}$) developed by \cite{wang2011weighted}, whose accuracy has not been systematically investigated.

The formulation of the grand potential energy density functional \eqref{grandpotential} is completed with the hard-wall component of the external potential, which away from the pore extremities is a function of $y$ alone,
  \begin{equation}\label{confiningpotential1} U_i^{\textrm{HW}}(y) =   \begin{cases} \displaystyle
  0  & \quad \text{if }  R_i \leq y \leq {L}-R_i, \\
    \infty  & \quad \text{otherwise.}
  \end{cases}
  \end{equation} 
This enforces non-overlapping of the pore wall and the spherical ions, with the density profiles of the latter dropping to zero within one radius of the wall.

\subsection{Asymptotic model reduction}

We consider the DDFT equation \eqref{DDFT3D} in 2D from here onwards, assuming uniform densities in the infinite depth dimension. The focus of the present work is the case of thin slit pores, for which we introduce a scale separation parameter $\epsilon = {L}/{H} \ll 1$. Additionally, we are only concerned with dynamics on the timescales of supercapacitor applications (with typical frequencies ranging from 0.001 to 1000 Hz) which are significantly longer than the nanosecond diffusion times of ionic liquids (see parameter sets below). Since a full asymptotic derivation of the model valid in the limit as $\epsilon \rightarrow 0$ is presented in \cite{aslyamov2020relation}, we provide an informal derivation following the discussions in \cite{chacon2014asymptotic} and \cite{narski2014asymptotic} for the related problem of anisotropic diffusion equations.

With the auxiliary pore length variable $\xi = \epsilon x$ and slow timescale $\tau = \epsilon^2 t$, we obtain the rescaled DDFT equation
\begin{equation}\label{DDFT2Dnondimrescale}
 \frac{\partial \rho_i}{\partial \tau}  =   \beta D_i \frac{\partial}{\partial \xi} \left( \rho_i  \frac{\partial \mu_i}{\partial \xi} \right) +  \frac{1}{\epsilon^2} \beta D_i \frac{\partial}{\partial y} \left( \rho_i  \frac{\partial \mu_i}{\partial y} \right),
\end{equation}
where, in $(\xi,y)$ coordinates, the pore is a square of side ${L}$. Since we do not wish to resolve the dynamics on the fast timescale, we may formally set $\epsilon = 0$ in the above equation to obtain
\begin{equation}\label{leadingorder}
0 =  \beta D_i \frac{\partial}{\partial y} \left( \rho_i  \frac{\partial \mu_i}{\partial y} \right).
\end{equation}
This is equivalent to the statement that, to leading order, the densities satisfy a 1D equilibrium DFT system in the pore width defined by the Euler--Lagrange equations \eqref{EL0} with constant, yet undetermined, chemical potentials $\mu_i \equiv \mu_i(x,t)$. It is important to note that, while the ion densities satisfy a 1D equilibrium problem, they are not necessarily \emph{in equilibrium} with the bulk phase external to the pore.

Equation \eqref{leadingorder} is a constraint that neither fully specifies an equilibrium problem at each $x$ nor informs the time-evolution of the system; to provide the correct well-posed limit model it must be allied with the appropriate integral equation \citep{narski2014asymptotic}. The latter is obtained by integrating the DDFT equation in $y$ and applying the no-flux boundary conditions, annihilating the singular term in \eqref{DDFT2Dnondimrescale} for any $\epsilon$. Taking the limit $\epsilon \rightarrow 0$, we may use the fact that the $\mu_i$ are independent of $y$, and the integral equation becomes, in $(x,t)$ variables,
\begin{equation} \label{weakform1}\frac{\partial N_i}{\partial t}  =   \beta D_i \frac{\partial}{\partial x} \left( N_i  \frac{\partial \mu_i}{\partial x} \right),
\end{equation}
where we define the particle numbers (per unit of pore width and depth)
\begin{equation}\label{particlenumberdefn}N_i(x,t) = \int_{0}^{L} \rho_i(x,y,t) \; \mathrm{d}y.\end{equation}
Equation \eqref{weakform1} describes the evolution of the particle numbers and is the same as the limit equation derived by \cite{aslyamov2020relation}. Accordingly, we solve a canonical 1D equilibrium problem at each $x$ for the ion densities as a function of $y$, where the particle numbers \eqref{particlenumberdefn} are prescribed. This is as opposed to solving a grand canonical problem where the chemical potentials are given. The numerical approach is discussed in Appendix \ref{numericsequilibrium}. As an output of this equilibrium problem, we obtain the chemical potentials $\mu_i$ at each $x$, gradients of which are required to evolve the particle numbers in time according to \eqref{weakform1}. There exist DFTs specifically for canonical calculations with a small integer number of particles \citep{de2014full}, e.g.~hard spheres in a closed cavity. However, since our slit pore extends infinitely in the direction perpendicular to the $x$--$y$ plane, we can perform a thermodynamic limit, and use DFTs derived originally in the grand canonical framework with \eqref{particlenumberdefn} as an additional constraint.

The 3D DFT components discussed above must be restricted to a planar 1D geometry. The 1D weight functions for the hard sphere terms can be found in \cite{roth2010fundamental}, while the expressions for the correlations according to the BFD and fMSA theories can be found in \cite{gillespie2002coupling} and \cite{roth2016shells}, respectively. For the electrostatics, the left hand side of the Poisson equation \eqref{Poisson0} becomes $\epsilon_0\epsilon_{\textrm{r}} \psi_{yy}$. Using Gauss' law, we can compute the surface charge density along the long pore walls as a function of time,
 \begin{equation}\label{NeumannPoissonnonsym1} \psi_y|_{y=0} = - \frac{\mathcal{Q}_0(x,t)}{\epsilon_0\epsilon_{\textrm{r}}}, \qquad \psi_y|_{y={L}} = + \frac{\mathcal{Q}_L(x,t)}{\epsilon_0\epsilon_{\textrm{r}}}.\end{equation}
 Integrating the Poisson equation over the full pore width, $y\in [0,L]$, and applying \eqref{NeumannPoissonnonsym1} gives the instantaneous charge neutrality expression for a fixed-$x$ slice of the pore,
\begin{equation}\label{integralequationSCD1}\mathcal{Q} = \mathcal{Q}_0 + \mathcal{Q}_L = -  {\textrm{e}} \sum_i z_i N_i,\end{equation}
where $\mathcal{Q}(x,t)$ denotes the sum of the local surface charge densities on the pore walls.

When the entire system is at equilibrium, the chemical potentials in the pore equal those in the bulk and $\mathcal{Q}$ is independent of $x$ and $t$. We can use \eqref{integralequationSCD1} to compute the the zero-frequency differential capacitance, $\mathrm{d}\mathcal{Q}/\mathrm{d}\mathcal{V}$ (denoting the rate of change in $\mathcal{V}$ for fixed $\mu_i$), and the integral capacitance, $\mathcal{Q}/(\mathcal{V}-\mathcal{V}_0)$, where $\mathcal{V}_0$ is the potential of zero charge (i.e.~$\mathcal{Q}(\mathcal{V}_0) = 0$).
The electrostatic component of the external potential, $U_i^{\textrm{ES}}$ (omitted above for brevity), is a function of the surface charge density, not the surface potential \citep{valisko2018systematic}; according to the uniqueness theorem of \cite{evans1979nature}, fixing $(\mathcal{Q}_0,\mathcal{Q}_L)$ uniquely determines the equilibrium densities in a planar geometry. However, the same cannot be said when prescribing surface potentials, and the surface charge--potential relation is known to be non-monotonic for certain parameters (which would provide more than one distinct equilibria for some values of the wall potential). For this reason, it is not ensured that $\mathcal{Q}_0=\mathcal{Q}_L$, or equivalently that the density profiles are symmetric about the pore centreline, $y = {L}/2$. Accordingly, all equilibrium profiles are solved for in a full pore, rather than a half pore with $\psi_y|_{y = {L}/2} =0$ (as is common in the literature), however, we did not observe any symmetry breaking or multiplicity of equilibria for the parameters considered here.

At the pore entrance and cap, the asymptotics break down and spatially 2D density profiles are expected; the full resolution of these regions is beyond the scope of the present study. At the entrance we fix the chemical potentials equal to those in the bulk,
\begin{equation} \label{entranceboundary} \mu_i = \mu_i^{\textrm{b}} \quad \textrm{at} \quad x = 0. \end{equation}
The particle numbers at $x=0$ are then computed from solving the grand canonical problem for 1D DFT equilibria (these will be time-dependent as the surface potential varies). 
At the pore cap, we ignore electrostatic effects due to the charged bounding wall and simply implement a no-flux condition,
\begin{equation} \label{capboundary} \frac{\partial \mu_i}{\partial x} = 0 \quad \textrm{at} \quad x = H.\end{equation}


\section{Impedance response analysis\label{ImpRes}}


\subsection{Semi-analytical approach}

Following a similar route to that outlined in \cite{huang2020review}, we construct an expression for the frequency response of the limit model which incorporates numerically computable sensitivities. The base state for the following linear analysis is a pore at equilibrium with the bulk phase ($\mu_i = \mu_i^{\textrm{b}}$), for which the equilibrium particle numbers are $\overline{N}_i$ given a potential $\overline{\mathcal{V}}$ applied on the pore walls. While the above theory is valid for general many-component mixtures of ionic liquids, we study two-component ionic liquids exclusively in this work. Variables and constants corresponding to the cation are indicated with a subscript $+$, while anion quantities have a subscript $-$. 

To mimic a potentiostatic impedance response experiment at the pore scale, we prescribe a voltage of the form $\mathcal{V}(t) = \overline{\mathcal{V}} + \Delta \mathcal{V}(t),$
where we use $\Delta$ to denote perturbation variables. Writing $\mu_i \equiv \mu_i(N_+,N_-,\mathcal{V})$, the chemical potential in the slit has the Taylor expansion
\begin{equation}
\mu_i = \mu_i^{\textrm{b}} + \frac{ \partial \mu_i}{\partial N_+} \Delta N_+ + \frac{ \partial \mu_i}{\partial N_-} \Delta N_-  + \frac{ \partial \mu_i}{\partial \mathcal{V}} \Delta \mathcal{V} + O(\Delta^2),
\end{equation}
where the partial derivatives are evaluated at the base state. We have the analytical fact that $\partial \mu_i/\partial \mathcal{V} = {\textrm{e}} z_i $, but the sensitivities with respect to the particle numbers must be computed numerically in general (these are calculated by solving the canonical equilibrium problem, as discussed in Appendix \ref{numericsequilibrium}, with perturbed particle numbers). The linearisation of \eqref{weakform1} about the equilibrium state is
\begin{equation}\label{linear1} \frac{\partial  \Delta N_i }{\partial t}  =   \beta D_{i} \overline{N}_i   \frac{\partial^2 }{\partial x^2} \left( \frac{ \partial \mu_i}{\partial N_+} \Delta N_+ + \frac{ \partial \mu_i}{\partial N_-} \Delta N_-  \right),
\end{equation}
with the boundary condition at the pore entrance \eqref{entranceboundary},
\begin{equation}\label{linear2}  \frac{ \partial \mu_i}{\partial N_+} \Delta N_+ + \frac{ \partial \mu_i}{\partial N_-} \Delta N_-  = -  {\textrm{e}} z_i \Delta \mathcal{V}   \quad \textrm{at} \quad x = 0, \end{equation}
and the no-flux condition \eqref{capboundary} at the pore cap,
\begin{equation}\label{linear3}  \frac{\partial }{\partial x} \left( \frac{ \partial \mu_i}{\partial N_+} \Delta N_+ + \frac{ \partial \mu_i}{\partial N_-} \Delta N_-  \right) = 0 \quad \textrm{at} \quad x = H. \end{equation}

With the solution of the above system, using \eqref{integralequationSCD1}, the perturbation to the local surface charge density, a function of $x$ and $t$, is calculated as
\begin{equation}\label{integralequationSCD1pert} \Delta \mathcal{Q} = - {\textrm{e}} z_+ \Delta N_+ - {\textrm{e}} z_- \Delta N_- .\end{equation}
Taking Fourier transforms in time (denoted with hats), we form the complex impedance per unit of pore depth and length (with units $\Omega$\,m$^2$),
\begin{equation} \label{Zdefn}
Z = \Delta \widehat{\mathcal{V}}\left( \frac{\textrm{i} \omega }{H} \int_0^H \Delta \widehat{\mathcal{Q}} \; \mathrm{d} x \right)^{-1},
\end{equation}
where $\omega = 2\pi f$ is the angular frequency for a temporal frequency $f$ (Hz). With this, we may obtain the frequency-dependent differential capacitance (in F\,m$^{-2}$) defined by
\begin{equation}\label{Cdiffdefn} C_{\textrm{diff}} = \frac{-1}{\omega \operatorname{Im}[Z]}. \end{equation}
The use of the average charge perturbation as opposed to the total charge perturbation in \eqref{Zdefn} is so that the differential capacitance \eqref{Cdiffdefn} is comparable with results for spatially 1D studies of slit pores, with the property that 
\begin{equation}\label{Cdifflimit} C_{\textrm{diff}} \xrightarrow{{\omega \rightarrow 0}}  \frac{ \mathrm{d}\mathcal{Q}}{\mathrm{d}\mathcal{V}}.\end{equation}

\subsubsection{Single ion species}

If one of the ion species is not present the pore, the equations for the perturbations (\ref{linear1},\ref{linear2},\ref{linear3}) are significantly reduced. For example, if $L < \sigma_+$, we have $\overline{N}_+ = 0$ and it can be surmised that $\Delta N_+ = 0$. The problem that remains for $ \Delta N_-$ is
\begin{equation} \frac{\partial  \Delta N_- }{\partial t}  =   \beta D_{-} \overline{N}_-   \frac{ \partial \mu_-}{\partial N_-}  \frac{\partial^2 \Delta N_- }{\partial x^2}  ,
\end{equation}
with
\begin{equation} \frac{ \partial \mu_-}{\partial N_-} \Delta N_- = -  {\textrm{e}} z_- \Delta \mathcal{V}   \quad \textrm{at} \quad x = 0, \qquad \frac{\partial \Delta N_-}{\partial x}   = 0 \quad \textrm{at} \quad x = H. \end{equation}
In frequency space, the solution is given by
\begin{equation}\Delta \widehat{N}_- = \frac{ {\textrm{e}} z_- }{(\partial \mu_-/\partial N_-)} \left[ \tanh(\sqrt{\textrm{i} \omega \lambda/\beta} H) \sinh(\sqrt{\textrm{i} \omega \lambda/\beta} x) - \cosh(\sqrt{\textrm{i} \omega \lambda/\beta} x)  \right]\Delta \widehat{\mathcal{V}},\end{equation}
where $\lambda = 1/ D_{-} \overline{N}_- (\partial \mu_-/\partial N_-)$. With the substitution of this expression and $\Delta \widehat{\mathcal{Q}} = - {\textrm{e}} z_- \Delta \widehat{N}_-$ into \eqref{Zdefn}, we obtain precisely the FSW impedance \eqref{deLeviemodel} with parameters
\begin{equation}\label{singleionexact}\mathcal{Z} = \frac{H^2}{ {\textrm{e}}^2 z_-^2 \beta D_{-} \overline{N}_- }, 
\qquad  \tau = \frac{H^2}{\beta D_{-} \overline{N}_- (\partial \mu_-/\partial N_-)}.\end{equation}
Using the zero frequency limit of the FSW element from \eqref{deLeviemodellimits} with (\ref{Cdiffdefn},\ref{Cdifflimit},\ref{singleionexact}) gives
\begin{equation}\label{singleiondQdv}\frac{\mathrm{d}\mathcal{Q}}{\mathrm{d}\mathcal{V}} = \frac{\tau}{\mathcal{Z}} = \frac{{\textrm{e}}^2 z_-^2 }{ \partial \mu_-/\partial N_-}.\end{equation}
This relation can also be obtained from $\mathcal{Q} = - {\textrm{e}} z_- N_-$ and the cyclic chain rule for partial derivatives (note that $\mathrm{d}\mathcal{Q}/\mathrm{d}\mathcal{V}$ denotes the derivative of $\mathcal{Q}$ respect to $\mathcal{V}$ with chemical potentials held fixed).

\subsubsection{General case}

In the situation that both ion species are present in the pore, we again take a Fourier transform in time of (\ref{linear1},\ref{linear2},\ref{linear3}) and work with the new variables
\begin{equation}\label{Ahatdefn} \widehat{A}_i =  \frac{ \partial \mu_i}{\partial N_+} \Delta \widehat{N}_+ + \frac{ \partial \mu_i}{\partial N_-} \Delta \widehat{N}_- ,
\end{equation}
with which we can write the system in the form
\begin{equation}\frac{\partial^2  }{\partial x^2} 
\bm{\widehat{A}} = \frac{\textrm{i} \omega }{\beta} M \bm{\widehat{A}} , \qquad \widehat{A}_i   = -  {\textrm{e}} z_i \Delta \widehat{\mathcal{V}}   \quad \textrm{at} \quad x = 0, \qquad \frac{\partial \bm{\widehat{A}} }{\partial x}  = 0 \quad \textrm{at} \quad x = H,\end{equation}
where the matrix $M$ is
\begin{equation}
M = \frac{1 }{ |J|  } 
\begin{pmatrix}
 (\partial \mu_-/\partial N_-) /D_{+} \overline{N}_+ & -  (\partial \mu_+/\partial N_-) /D_{+} \overline{N}_+  \\
-  (\partial \mu_-/\partial N_+)/D_{-} \overline{N}_- & (\partial \mu_+/\partial N_+) /D_{-} \overline{N}_-  
\end{pmatrix}, \quad |J| = \frac{ \partial \mu_+}{\partial N_+}   \frac{ \partial \mu_-}{\partial N_-} -  \frac{ \partial \mu_+}{\partial N_-} \frac{ \partial \mu_-}{\partial N_+}.
\end{equation}
If $\lambda_1$ and $\lambda_2$ denote the eigenvalues of $M$, corresponding to eigenvectors $\bm{v}_1 = (v_1,1)^{T}$ and $\bm{v}_2 = (v_2,1)^{T}$, respectively, the solution to this system is
\begin{align}
\frac{\widehat{A}_+}{{\textrm{e}} \Delta \widehat{\mathcal{V}}} = & \; v_1 a_1 \left[\tanh(\sqrt{\textrm{i} \omega \lambda_1 /\beta} H)  \sinh(\sqrt{\textrm{i} \omega \lambda_1 /\beta} x)  -   \cosh(\sqrt{\textrm{i} \omega \lambda_1 /\beta} x) \right] \nonumber \\
& + v_2 a_2\left[  \tanh(\sqrt{\textrm{i} \omega \lambda_2 /\beta} H) \sinh(\sqrt{\textrm{i} \omega \lambda_2 /\beta} x) - \cosh(\sqrt{\textrm{i} \omega \lambda_2 /\beta} x) \right] ,\\
\frac{\widehat{A}_-}{ {\textrm{e}} \Delta \widehat{\mathcal{V}}} = & \; a_1 \left[  \tanh(\sqrt{\textrm{i} \omega \lambda_1 /\beta} H)  \sinh(\sqrt{\textrm{i} \omega \lambda_1 /\beta} x) - \cosh(\sqrt{\textrm{i} \omega \lambda_1 /\beta} x) \right] \nonumber \\
& + a_2 \left[ \tanh(\sqrt{\textrm{i} \omega \lambda_2 /\beta} H) \sinh(\sqrt{\textrm{i} \omega \lambda_2 /\beta} x)  -  \cosh(\sqrt{\textrm{i} \omega \lambda_2 /\beta} x) \right],
\end{align}
where $(a_1,a_2)^T$ satisfies
\begin{equation}\begin{pmatrix} 
v_1 & v_2 \\
1 & 1
\end{pmatrix}
\begin{pmatrix} 
 a_1 \\
 a_2
\end{pmatrix}
= 
\begin{pmatrix} 
 z_+ \\
  z_-
\end{pmatrix}.
\end{equation}
Defining time constants $\tau_i = H^2  \lambda_i /\beta$, the complex impedance \eqref{Zdefn} can be expressed as
\begin{align}
 \frac{ H^2 }{{\textrm{e}}^2 \beta} \; Z^{-1}  = & \;  z_+ D_{+} \overline{N}_+ \left[ v_1  a_1  \sqrt{\textrm{i} \omega \tau_1}   \tanh(\sqrt{\textrm{i} \omega \tau_1} ) + v_2  a_2 \sqrt{\textrm{i} \omega \tau_2}  \tanh(\sqrt{\textrm{i} \omega \tau_2}) \right]  \nonumber \\
& +  z_-  D_{-} \overline{N}_- \left[ a_1 \sqrt{\textrm{i} \omega \tau_1}  \tanh(\sqrt{\textrm{i} \omega \tau_1}) + a_2 \sqrt{\textrm{i} \omega \tau_2}  \tanh(\sqrt{\textrm{i} \omega \tau_2}) \right]. \label{fullZ}
\end{align}

For small frequencies, the complex impedance behaves according to
\begin{align}
  \frac{ H^2 }{{\textrm{e}}^2 \beta} \; Z^{-1}  \xrightarrow{{\omega \rightarrow 0}} & \;  \frac{\omega^2 }{3} \left(  z_+ D_{+} \overline{N}_+  \left[  v_1  a_1 \tau_1^2 +  v_2  a_2 \tau_2^2 \right]  
 +  z_-  D_{-} \overline{N}_-  \left[  a_1 \tau_1^2 +  a_2 \tau_2^2 \right] \right)\nonumber \\
& 
 +   \omega \left( z_+ D_{+} \overline{N}_+ \left[ v_1  a_1 \tau_1 + v_2  a_2 \tau_2 \right]  +  z_-  D_{-} \overline{N}_-  \left[ a_1 \tau_1  + a_2 \tau_2  \right] \right) \textrm{i} . \label{smallomegaanalytical}
\end{align}
The asymptotic behaviour \eqref{smallomegaanalytical} fully prescribes an ``effective" FSW element \eqref{deLeviemodel}, valid for small frequencies, with parameters denoted $\mathcal{Z}_0$ and $\tau_{0}$; these are found by substituting $Z^{-1} \approx \mathcal{Z}_0^{-1}(\textrm{i} \omega \tau_0 + \omega^2 \tau_0^2/3)$ into \eqref{smallomegaanalytical}. The zero frequency limit \eqref{smallomegaanalytical} can also be used to compute a relation between the chemical potential sensitivities and $\mathrm{d}\mathcal{Q}/\mathrm{d}\mathcal{V}$, although more convoluted than \eqref{singleiondQdv}, and we may also obtain the asymptote of $\operatorname{Re}[Z]$,
\begin{equation}
\frac{3{\textrm{e}}^2 \beta }{  H^2} \operatorname{Re}[Z] \xrightarrow{{\omega \rightarrow 0}}  \frac{ z_+ D_{+} \overline{N}_+  \left[  v_1  a_1 \tau_1^2 +  v_2  a_2 \tau_2^2 \right]  
 +  z_-  D_{-} \overline{N}_-  \left[  a_1 \tau_1^2 +  a_2 \tau_2^2 \right] }{  \left( z_+ D_{+} \overline{N}_+ \left[ v_1  a_1 \tau_1 + v_2  a_2 \tau_2 \right]  +  z_-  D_{-} \overline{N}_-  \left[ a_1 \tau_1  + a_2 \tau_2  \right] \right)^2}.
\end{equation}
In the large $\omega$ limit, we have
\begin{equation}
 \frac{ (1-\textrm{i}) H^2 }{{\textrm{e}}^2 \beta \sqrt{2\omega }  } \; Z^{-1}  \xrightarrow{{\omega \rightarrow \infty}}  z_+ D_{+} \overline{N}_+ \left[ v_1  a_1  \sqrt{ \tau_1 } + v_2  a_2 \sqrt{ \tau_2 } \right] +  z_-  D_{-} \overline{N}_- \left[ a_1 \sqrt{ \tau_1 } + a_2 \sqrt{ \tau_2 } \right], \label{largeomegaanalytical}
\end{equation}
which fixes one of the two degrees of freedom required to determine an effective FSW element with parameters $(\mathcal{Z}_{\infty},\tau_{\infty})$. We can identify that the correct time constant is $\tau_{\infty} = \min\{\tau_1,\tau_2\}$ by including the exponentially small terms in the above expansion and in the corresponding FSW limit in \eqref{deLeviemodellimits}. In the case that one ion species is negligible in comparison to the other, e.g.~$\overline{N}_- \gg \overline{N}_+ > 0 $, both $\tau_0$ and $\tau_{\infty}$ can be approximated by the FSW time constant in \eqref{singleionexact}, involving a sensitivity with respect to the non-negligible species. We observed that $\tau_0$ diverged from the single species result for much smaller co-ion particle numbers than $\tau_{\infty}$ (which can remain close to \eqref{singleionexact} for particle numbers separated by an order of magnitude only).




\subsection{Time-dependent numerical simulations\label{timedepsection}}

We additionally perform time-dependent simulations of the limit model to validate the semi-analytical impedance expressions. This step validates the numerical scheme itself, which will be used for future investigations of ionic liquids in long nano-pores. The details of the finite-volume scheme, as well as its particular implementation for frequency response simulations, are described in Appendix \ref{numericstime}. The approach is similar to that utilised by \cite{aslyamov2020relation}, but we take care to use time-stepping schemes which preserve positivity of particle numbers (a particularly desirable property for simulating pore systems for which some of the ion species can become negligibly small) and use a non-uniform spatial grid to reduce numerical expense.

For our dynamic simulations, we initialise with a system at equilibrium and apply a sinusoidal voltage signal of amplitude ${\max_t|\Delta \mathcal{V}(t)|} = 2.5$ mV. The average surface charge density in the pore is computed by integrating \eqref{integralequationSCD1} in $x$ and dividing the result by $H$ (approximated with a rectangle rule for the discretised system). After each period, we estimate the complex impedance by fitting the magnitude, $|Z|$, and phase, $\arg(Z)$, with the input voltage signal and output average surface charge density (fast Fourier transforms can be used when the data is equispaced in time).

\section{Numerical results\label{numresults}}

We present results for two parameter sets based on the monovalent ($z_+ = - z_- = 1$) ionic liquids [EMIm][TFSI] and [EMIm][BF$_4$] at 300K, summarised in Table \ref{physproptable}:
\begin{table}[H]
\begin{center}
\begin{tabular}{ |c|c|c|c| }
 \hline 
Ionic liquid & I -- [EMIm][TFSI] & II -- [EMIm][BF$_4$] \\ 
 \hline
$\epsilon_{\textrm{r}}$ &  12  & 14.5  \\ 
$\sigma_+$ (nm) & 0.5 & 0.5  \\ 
$\sigma_-$ (nm) & 0.5 & 0.3  \\ 
$D_+$ ($\times 10^{-11}$ m$^2$s$^{-1}$) & 6.746 & 5.322 \\ 
$D_-$ ($\times 10^{-11}$ m$^2$s$^{-1}$) & 3.771 & 4.461  \\ 
$\sigma_+^2/D_+$ (ns) & 3.706 & 4.698 \\ 
 \hline
\end{tabular} 
\end{center}
\caption{Physical parameters based on [EMIm][TFSI] and [EMIm][BF$_4$].}\label{physproptable}
\end{table}
The relative permittivities (dielectric constants) given in the first row are the static (zero-frequency limit) bulk values obtained by dielectric relaxation spectroscopy \citep{huang2011static,hunger2009temperature}, and are assumed fixed for the model ionic liquids. The ion diameters in the second and third rows of Table \ref{physproptable} are commonly used approximations for the asymmetric molecules under consideration as spherical particles \citep{neal2017ion}, while the (constant) self-diffusion coefficients in fourth and fifth rows of Table \ref{physproptable} are obtained from the Vogel--Tammann--Fulcher equations (based on data from NMR experiments) provided by \cite{noda2001pulsed}. The characteristic diffusion timescale is given in the final row of Table \ref{physproptable}. The parameters for case I were chosen to exemplify an ionic liquid composed of ions with similar sizes (restricted primitive model) with widely differing diffusivities, whereas case II represents ionic liquids exhibiting strong size asymmetry. Bulk densities of the cation and anion are set equal to $2.4$ nm$^{-3}$ (3.985 mol/L), providing the non-dimensional reduced bulk densities $\sigma_+^{3} \rho_+^{\textrm{b}} = \sigma_+^{3} \rho_-^{\textrm{b}} = 0.3$. This is close to the typical value for the ionic liquids which inspired the above parameter sets, and has been considered in other works \citep{qingsurface}. It follows from the limit model (and the expressions for the complex impedance) that changing the pore length results in a simple rescaling of the impedance response, and so we fix $H = 100$ $\mu$m for our numerical results. All numerical calculations were performed in a non-dimensional framework (see Appendix \ref{numericstime}). The use of dimensional variables in the figures below allows the comparison of the two ionic liquids (of differing characteristic diffusion times) with ease.


We consider four different DFTs, the simplest of these being the DCA-PB model \citep{hoffmann2013ion,gillespie2015review} which is the incorporation of steric hard-wall interactions \eqref{confiningpotential1} into the standard PB model (i.e.~$F^{\textrm{HS}} = F^{\textrm{EC}} = 0$). Without hard-wall effects, contact densities are unrealistically large and the equilibrium problem becomes difficult to resolve numerically, so the standard PB model is not included in our study. We also consider the addition of (White Bear II) hard-sphere terms to the DCA-PB model, which we refer to as the HS-PB model. The two remaining DFTs include all these elements as well as electrostatic correlations, and are identified by the model used for the latter (either BFD or fMSA). Equilibrium density profiles for the four DFTs with ionic liquid I are shown in Figure \ref{paramsetIequilibria} in Appendix \ref{numericsequilibrium}.

\begin{figure}
\centering
\begin{subfigure}{2.8in}\caption{Resistance.}
\includegraphics[width=2.55in]{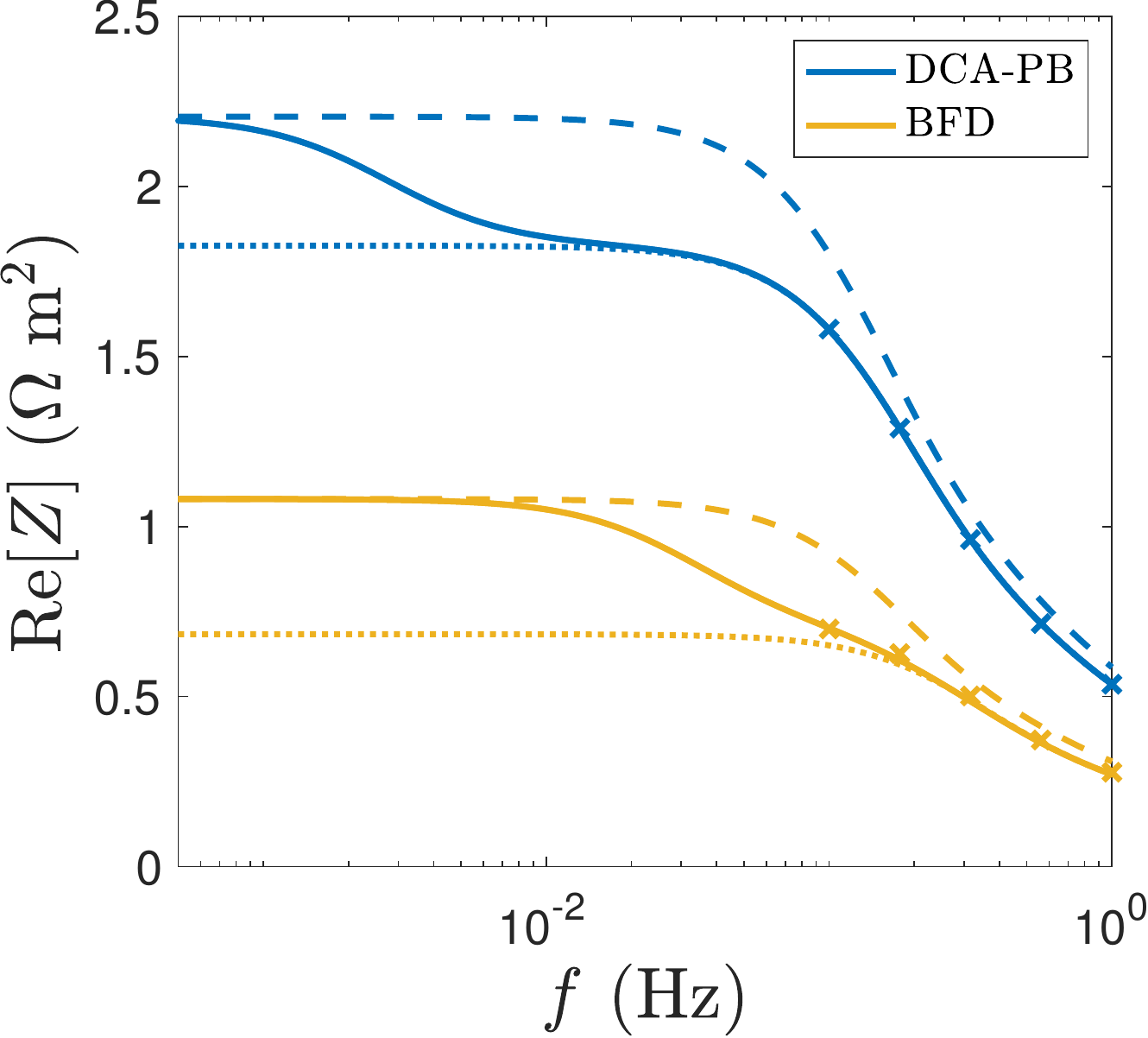}
\end{subfigure}
\begin{subfigure}{2.8in}\caption{Differential capacitance.}
\includegraphics[width=2.55in]{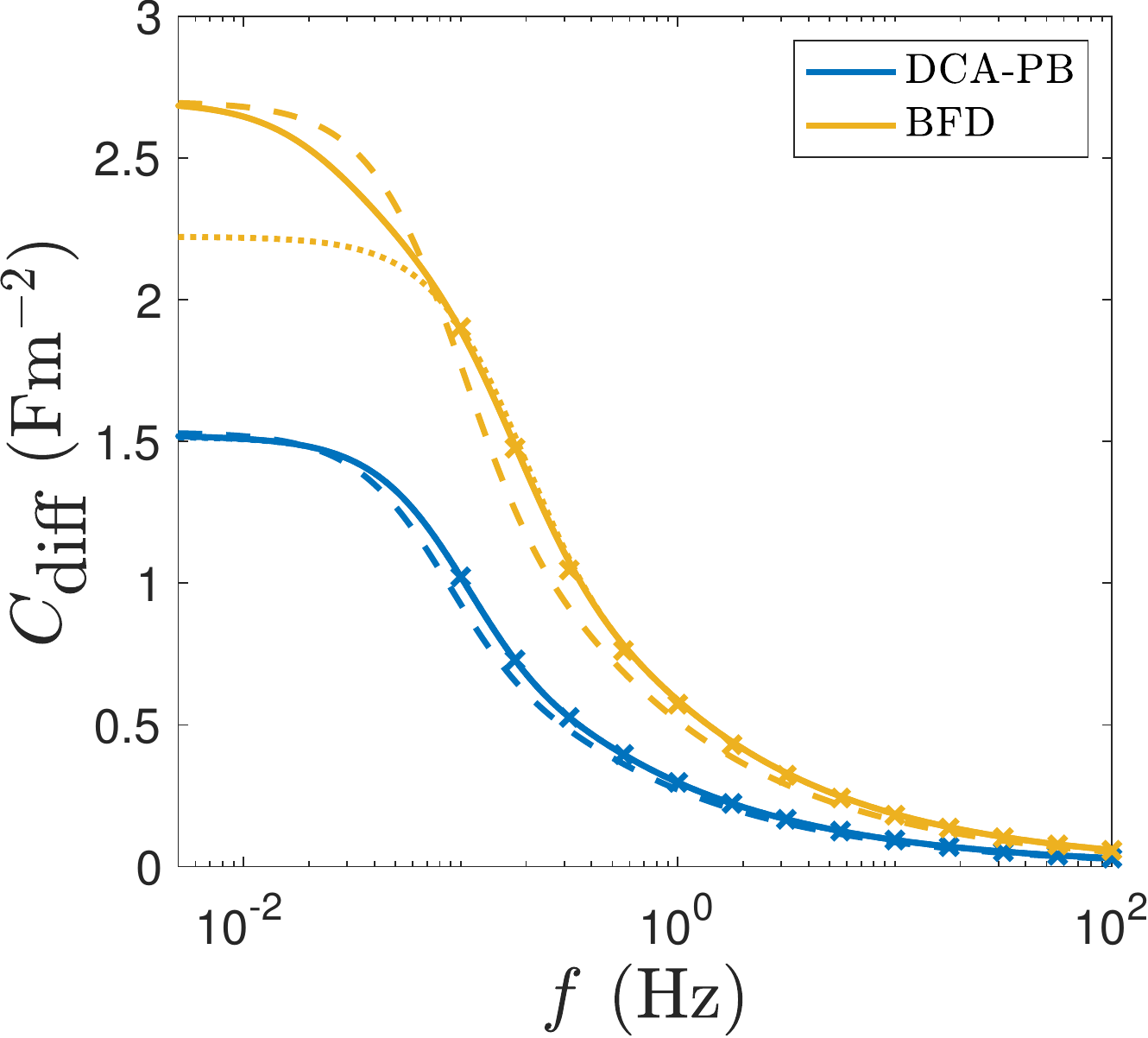}
\end{subfigure}
\caption{Impedance response for a pore of width $L = 2\sigma_+$ with the parameters of ionic liquid II and $\overline{\mathcal{V}} = 0.5$. Solid lines correspond to the full semi-analytical result \eqref{fullZ}, with the dotted and dashed lines indicating the high and low frequency effective FSW elements with parameters $(\mathcal{Z}_{\infty},\tau_{\infty})$ and $(\mathcal{Z}_{0},\tau_{0})$, respectively. The markers denote the results of time-dependent numerical simulations.}\label{BF_Cdiff_plots}
\end{figure}

In Figure \ref{BF_Cdiff_plots}, we plot the (a) resistance (real part of the complex impedance) and (b) differential capacitance \eqref{Cdiffdefn} against frequency for ionic liquid II with base voltage is $\overline{\mathcal{V}} = 0.5$ V. The solid line, computed from the full expression for the complex impedance \eqref{fullZ}, exhibits a double plateau in panel (a). This is expected for the impedance response of a system with two characteristic times ($\tau_1$, $\tau_2$). The dashed and dotted lines, denoting the results for the low and high frequency effective FSW elements, respectively, bound the solid line above and below. The BFD curves in panel (b) show a similar transition from the low to high frequency behaviour in the differential capacitance, whereas the asymptotic curves for DCA-PB are very close to the true differential capacitance across the full frequency range. The plots for HS-PB and fMSA are similar to those for DCA-PB and BFD shown in the figure, respectively, and so are not included. It is clear from Figure \ref{BF_Cdiff_plots} that the impedance response can deviate significantly from the typical FSW element obtained for a straight slit pore with a simple physical model. The markers in Figure \ref{BF_Cdiff_plots}, indicating the results of time-dependent numerical simulations, track the solid line very closely (it is too numerically expensive to simulate the system deeper into the low frequency regime). Beyond the results shown in Figure \ref{BF_Cdiff_plots}, we carried out time-dependent numerical simulations for each combination of DFT and ionic liquid, for $L/\sigma_+ = 2,4,6,8$, and a range of frequencies in the interval $0.1\; \textrm{Hz} < f < 100 \; \textrm{Hz}$. We found a similarly good agreement between the semi-analytical and fully numerical approaches in all cases (data not shown).

\begin{figure}
\centering
\begin{subfigure}{3in}\caption{High frequency response time (I).}
\includegraphics[width=2.75in]{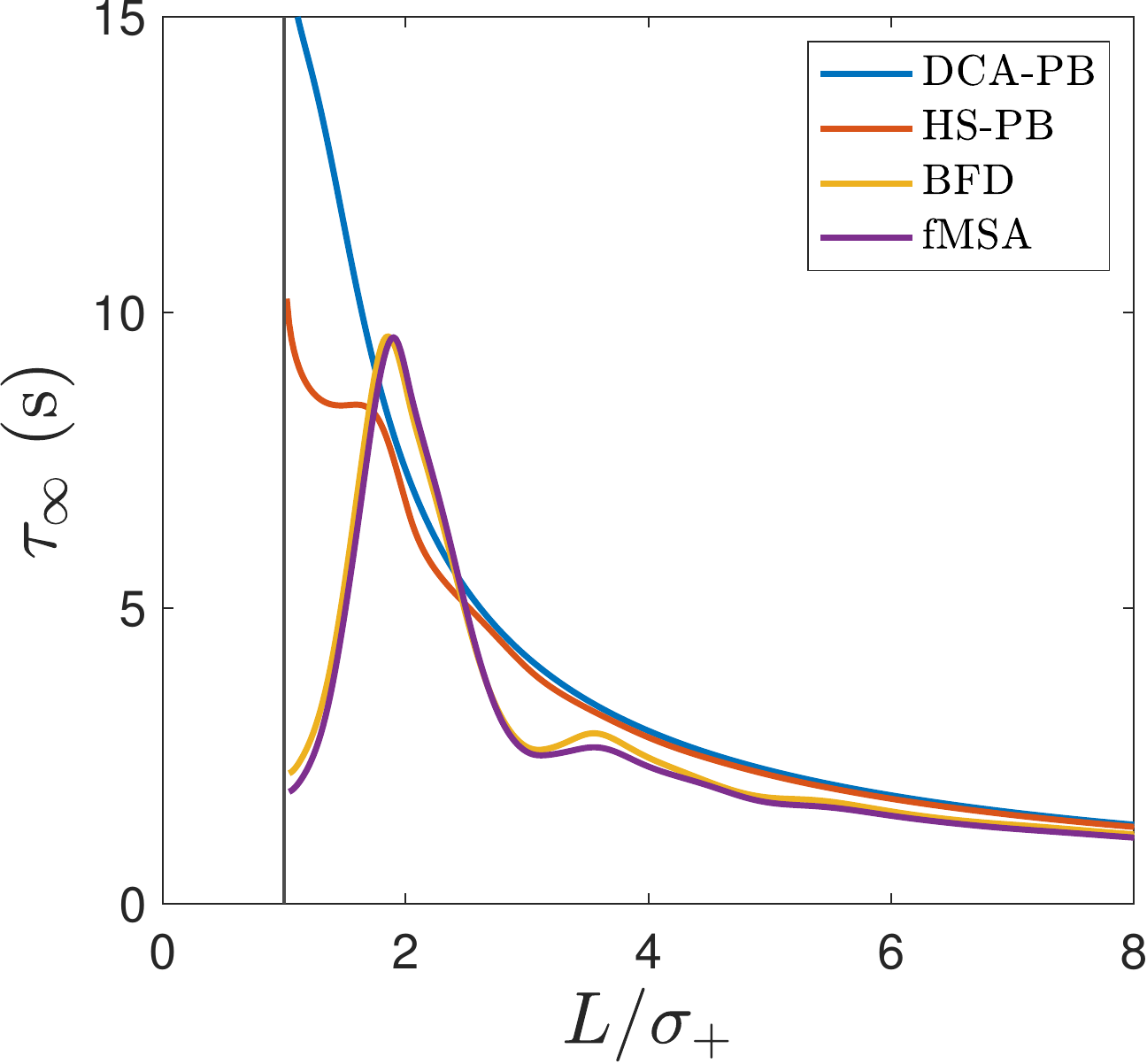}
\end{subfigure}
\begin{subfigure}{3in}\caption{High frequency response time (II).}
\includegraphics[width=2.75in]{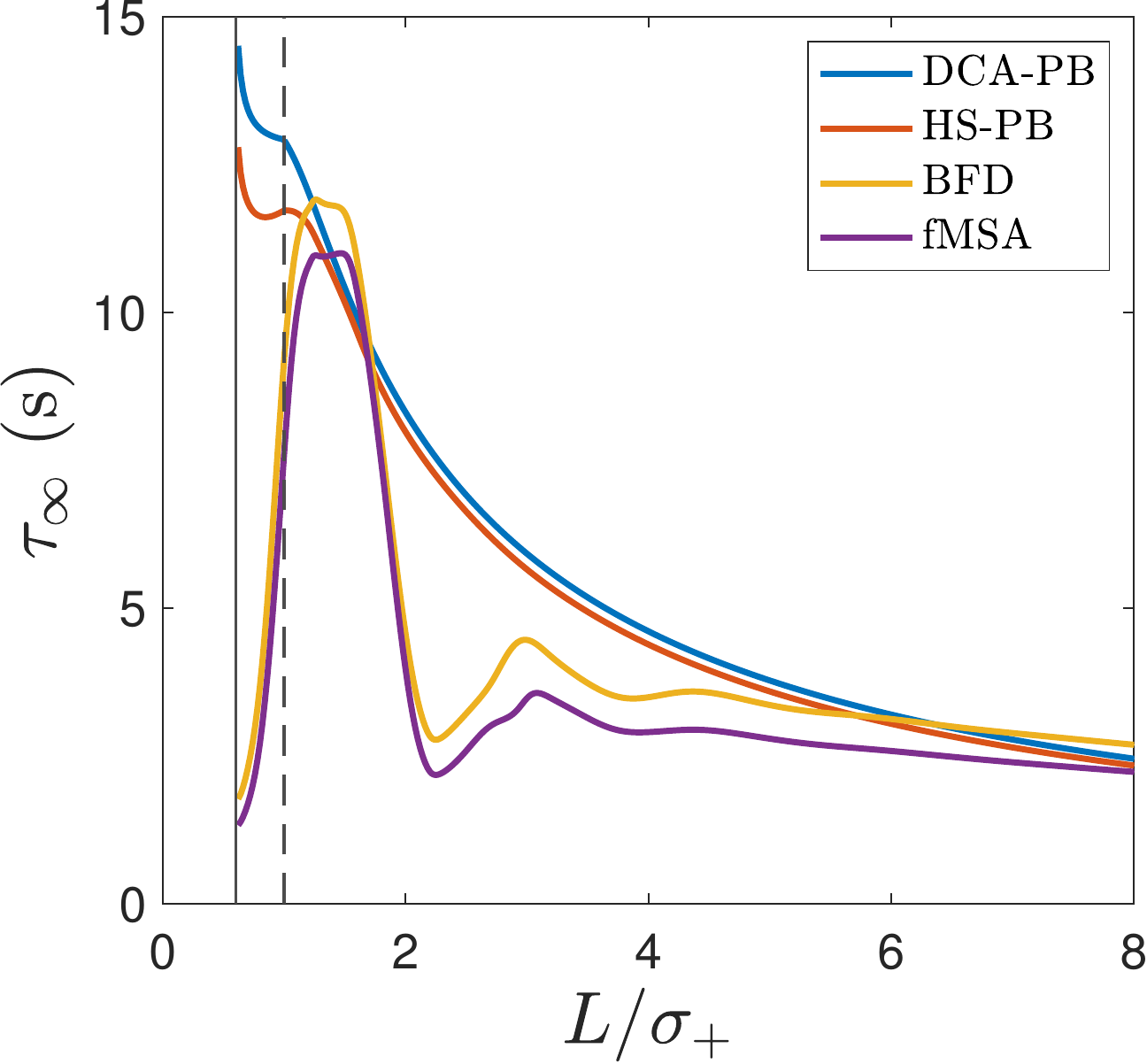}
\end{subfigure}
\centering
\begin{subfigure}{3in}\caption{Low frequency response time (I).}
\includegraphics[width=2.75in]{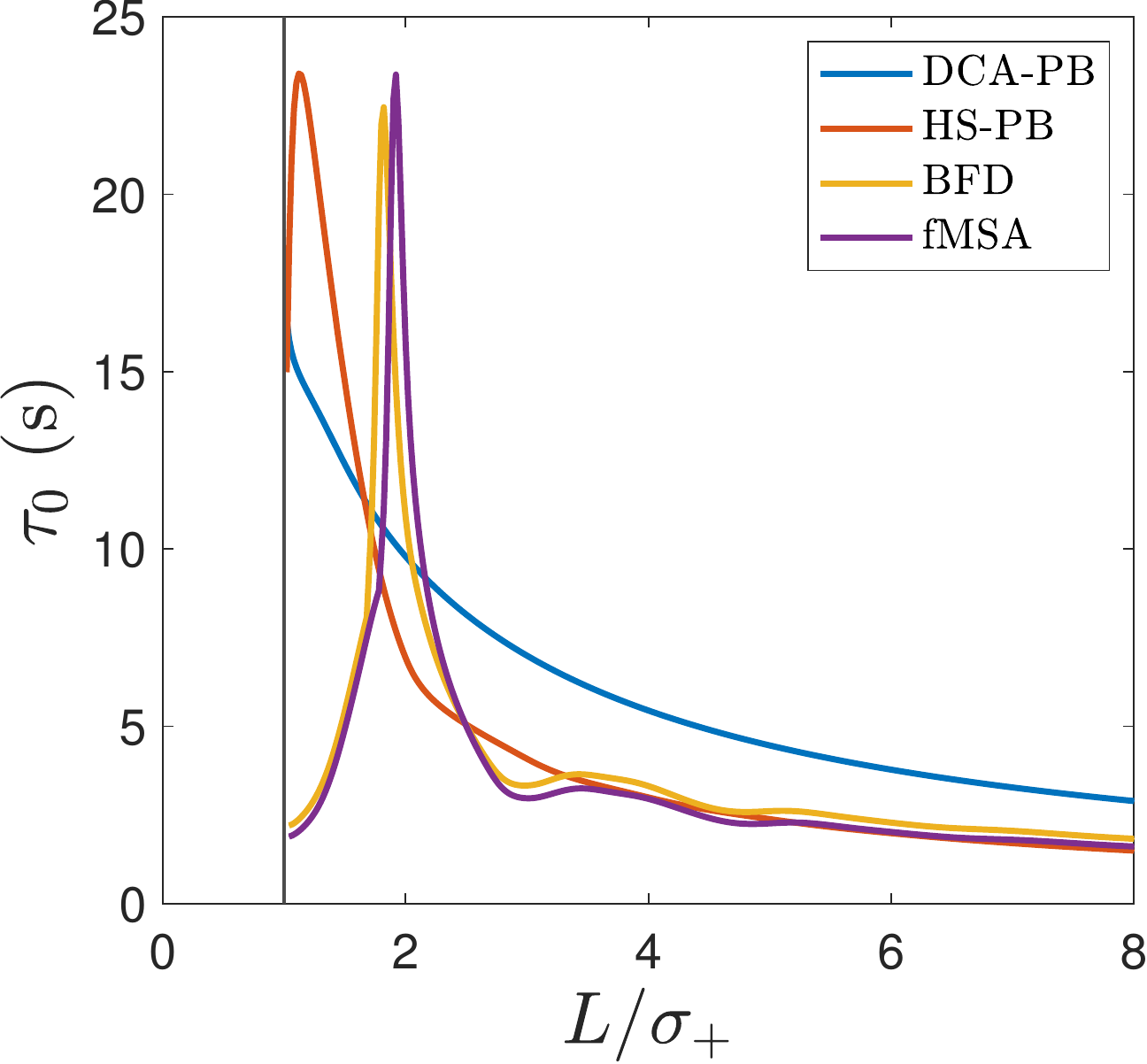}
\end{subfigure}
\begin{subfigure}{3in}\caption{Low frequency response time (II).}
\includegraphics[width=2.75in]{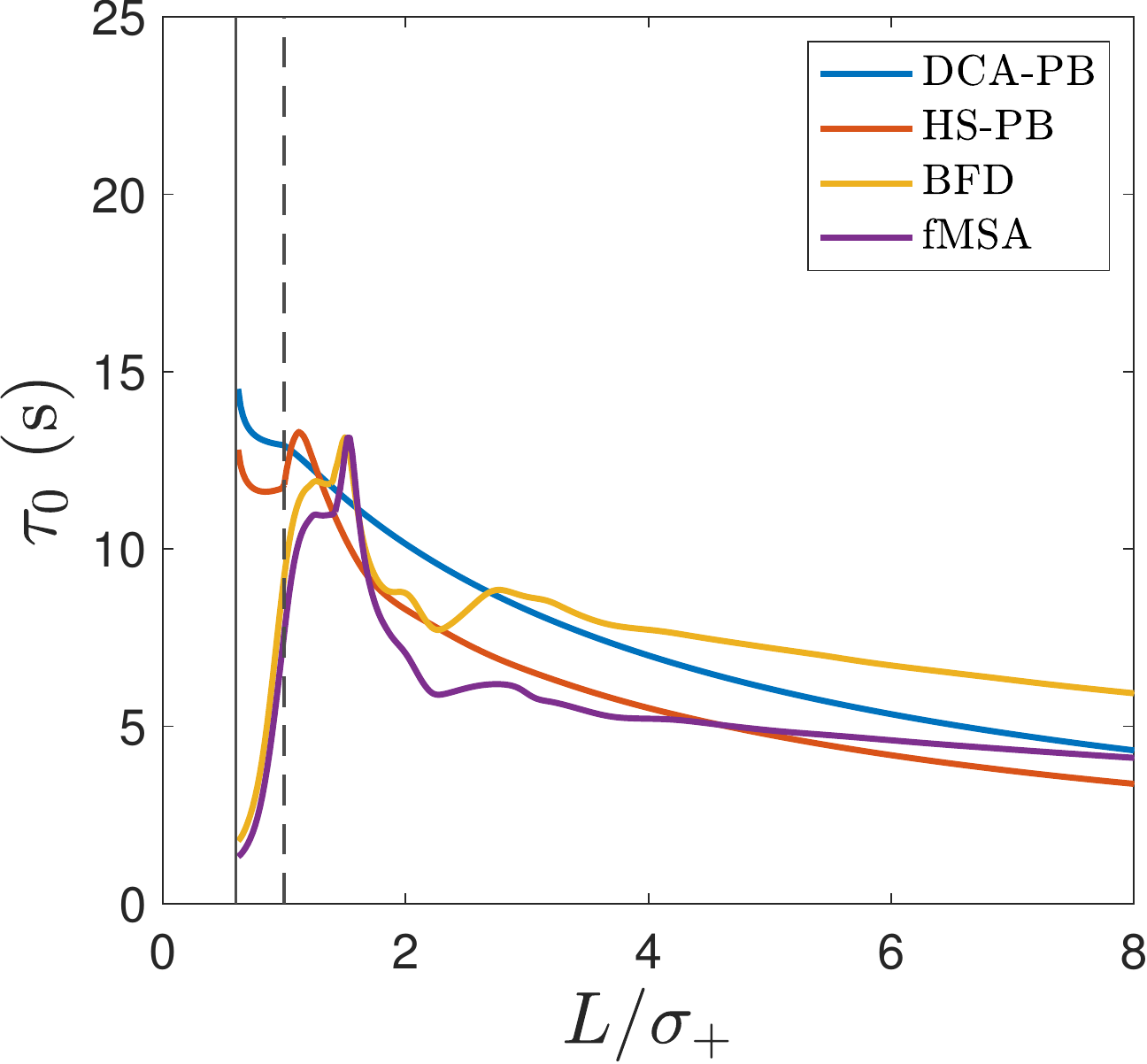}
\end{subfigure}
\caption{High (a,b) and low (c,d) frequency response time against pore width at $\overline{\mathcal{V}} = 0.5$ V. The results for ionic liquid I are shown in panels (a,c), with ionic liquid II in (b,d). The solid curves indicate the effective FSW time constants ($\tau_{\infty}$ and $\tau_{0}$) for each of the DFTs. The solid vertical lines correspond to the widths below which no ions can fit in the pore, with the dashed vertical lines in panels (b,d) indicating $L = \sigma_+$, below which only the smaller anion can enter the pore. For $\sigma_- < L < \sigma_+$, the curves in panels (b,d) are the exact FSW time constants for the single-ion system \eqref{singleionexact}.}\label{FSW_tau_plots}
\end{figure}

In Figure \ref{FSW_tau_plots}(a,b), we plot $\tau_{\infty}$ against pore width for each of the four DFTs, again with $\overline{\mathcal{V}} = 0.5$ V. The results for the size-symmetric ionic liquid I are shown in panel (a), with the size-asymmetric ionic liquid II in panel (b). The solid and dashed vertical lines in Figure \ref{FSW_tau_plots}(b) bound the region in which the larger cation cannot enter the slit pore; the value of $\tau_{\infty}$ here is exact and given in \eqref{singleionexact}. For the BFD and fMSA curves, the transition across the dashed line is smooth since $\overline{N}_+$ is negligible as $L \rightarrow \sigma_+$ (from above). For the DFTs without electrostatic correlations (DCA-PB and HS-PB), the response time decreases with pore width. When $L > 2 \sigma_+ $, the values of $\tau_{\infty}$ for these two theories agree quite closely (this is due to the dominance of electrostatic effects for our ionic liquid-inspired parameters, with a larger disparity found for $\epsilon_{\textrm{r}} = 40$, say). For the DFTs including screening effects, the value of $\tau_{\infty}$ when $L \rightarrow \sigma_-$ is relatively small, with a peak observed at $L = 2\sigma_+$ in panel (a), and a large $\tau_{\infty}$ regime found across $1.25 < L/\sigma_+ < 1.5$ in panel (b). There is good agreement between the results for BFD and fMSA, with a noticeable shift in $\tau_{\infty}$ for the size-asymmetric ionic liquid in panel (b) -- the shapes of the curves match well, however. In both panels (a) and (b), the high frequency response times for all DFTs converge to similar values for large pores ($L \gtrsim 6 \sigma_+$).

The corresponding plots for $\tau_0$ are shown in Figure \ref{FSW_tau_plots}(c,d). In panel (c), we see a significant deviation between the results for DCA-PB and HS-PB across all $L$, with the latter agreeing closely with BFD and fMSA for $L \gtrsim 4 \sigma_+$. For the size-asymmetric case in panel (d), we again see wide peaks in the time constant and a disagreement between the DFTs for large $L$. While the peaks in $\tau_{\infty}$ appear to be more significant for ionic liquid II, the situation is reversed for $\tau_0$. Much of the structure of $\tau_0$ for small $L$ is observed in $\mathrm{d}\mathcal{Q}/\mathrm{d}\mathcal{V}$ (as the latter determines one degree of freedom for the low frequency FSW element).


\begin{figure}
\centering
\begin{subfigure}{3in}\caption{$\overline{\mathcal{V}} \geq 0$ (I).}
\includegraphics[width=2.75in]{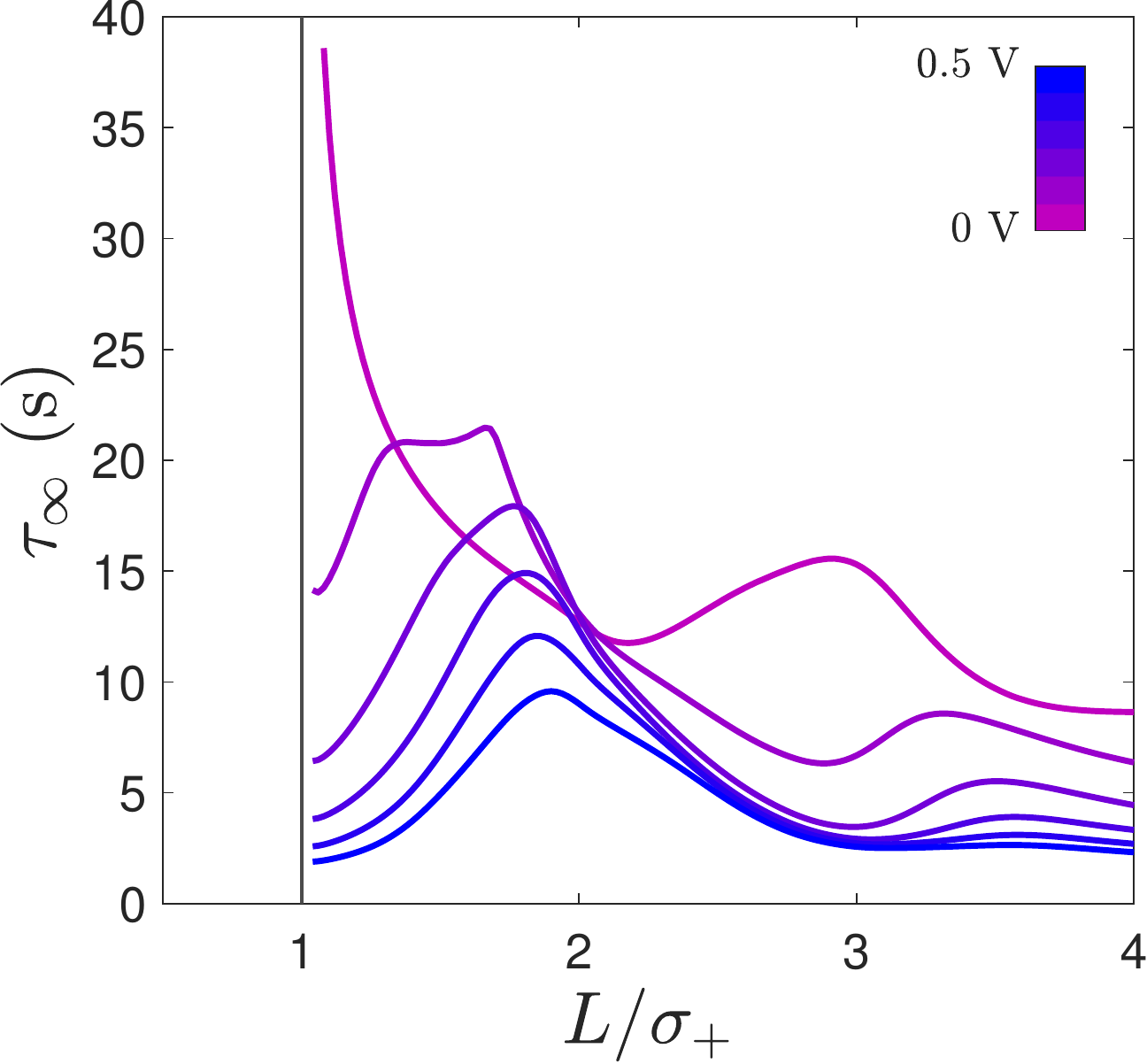}
\vspace{3mm}
\end{subfigure}
\begin{subfigure}{3in}\caption{$\overline{\mathcal{V}} \geq 0$ (II).}
\includegraphics[width=2.75in]{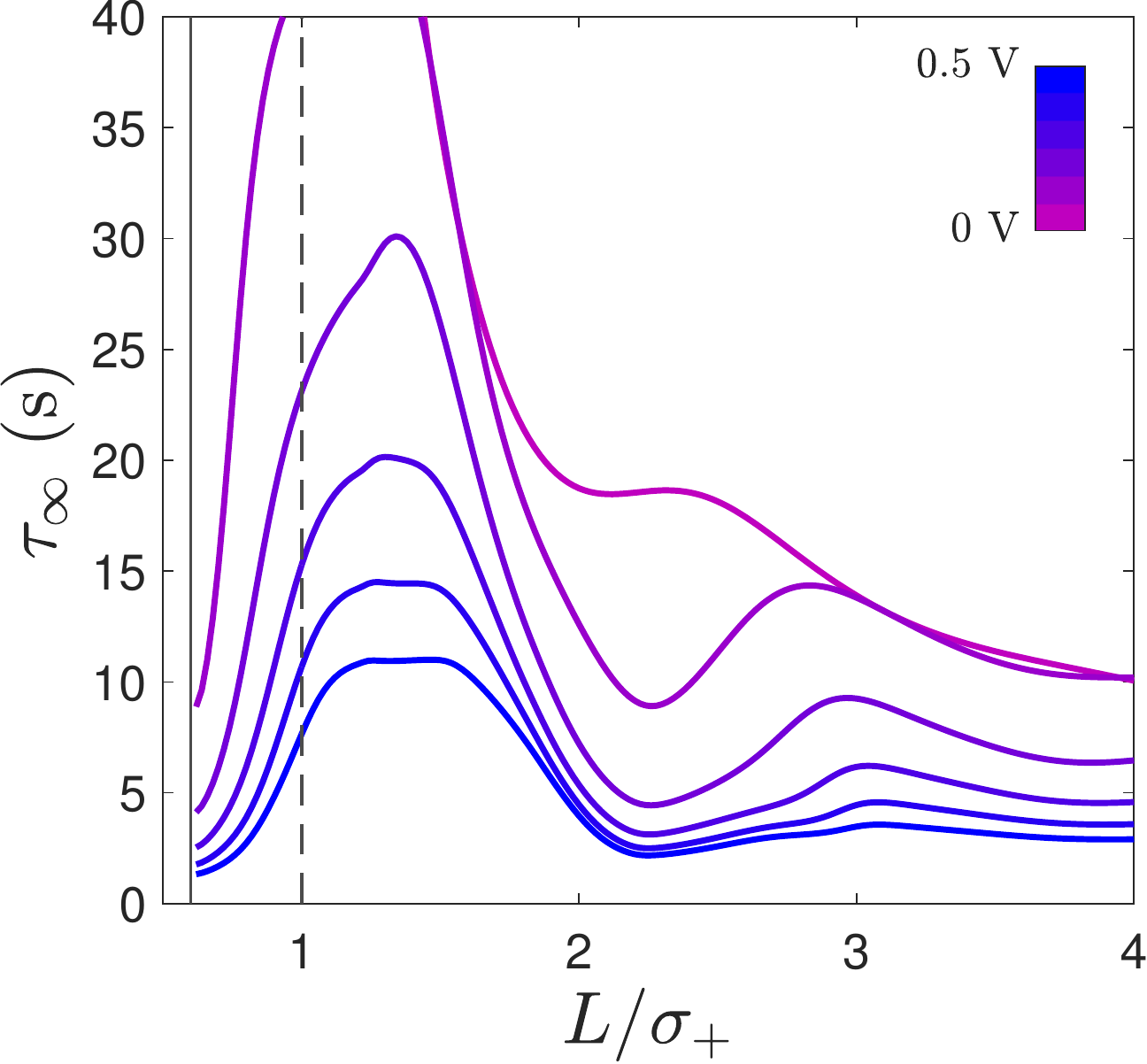}
\vspace{3mm}
\end{subfigure}
\begin{subfigure}{3in}\caption{$\overline{\mathcal{V}} \leq 0$ (I).}
\includegraphics[width=2.75in]{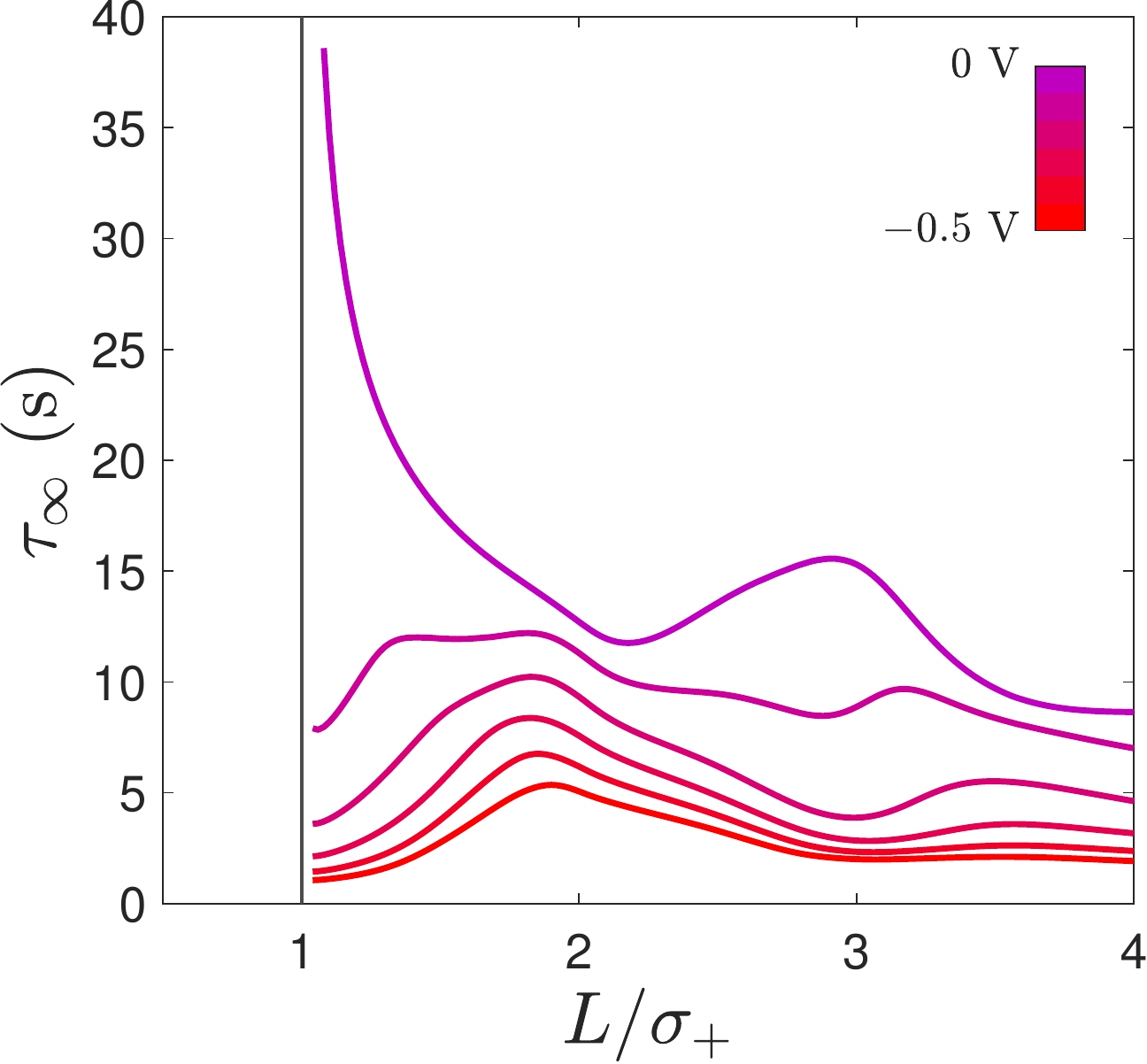}
\end{subfigure}
\begin{subfigure}{3in}\caption{$\overline{\mathcal{V}} \leq 0$ (II).}
\includegraphics[width=2.75in]{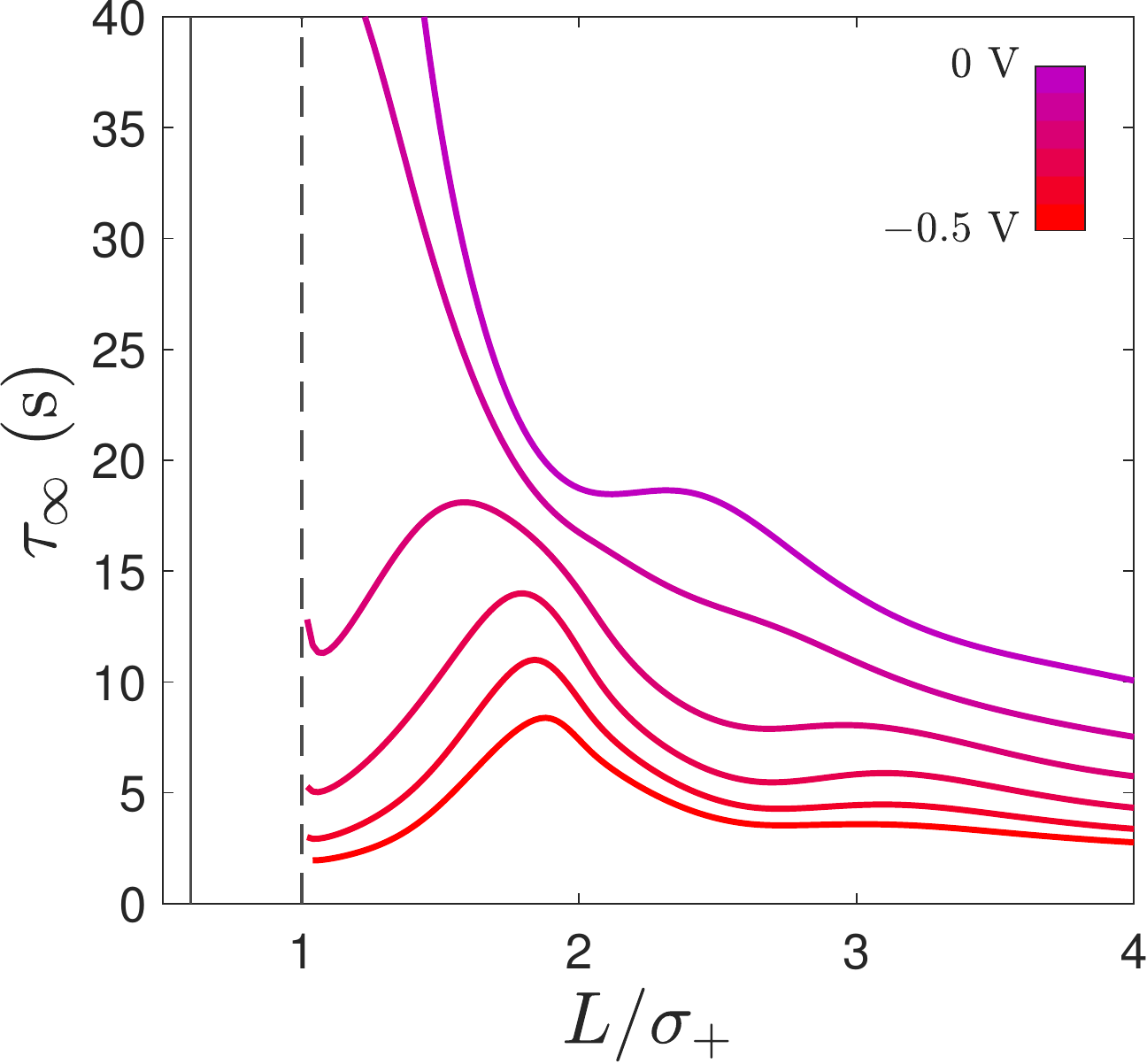}
\end{subfigure}
\caption{High frequency response time against pore width for a range of $\overline{\mathcal{V}}$ with fMSA for the DFT. Panels (a,b) show curves corresponding to $\overline{\mathcal{V}} = 0, 0.1, 0.2, 0.3, 0.4, 0.5$ V, with $-\overline{\mathcal{V}} =  0, 0.1, 0.2, 0.3, 0.4, 0.5$ V in panels (c,d). The left/right panels correspond to ionic liquid I/II, with the solid and dashed vertical lines as in Figure \ref{FSW_tau_plots}.}\label{FSW_Vscan_plots}
\end{figure}

For fMSA, we further investigated the dependence of the impedance response on the base voltage $\overline{\mathcal{V}}$, the results of which are presented in Figure \ref{FSW_Vscan_plots}. The top/bottom panel corresponds to positive/negative wall potentials, with the results for ionic liquids I and II shown in the left and right panels, respectively. The fMSA lines in Figure \ref{FSW_tau_plots} are the curves of smallest magnitude in panels (a,b) of Figure \ref{FSW_Vscan_plots}. For the size-symmetric ionic liquid in Figure \ref{FSW_Vscan_plots}(a,c), we see the large $\tau_{\infty}$ regime widen and shift from $L \approx 2 \sigma_+$ for $|\overline{\mathcal{V}}| = 0.5$ V towards $L \approx \sigma_+$ as $|\overline{\mathcal{V}}|$ is decreased to zero (potential of zero charge). A second peak of smaller magnitude shifts similarly, found at $L$ approximately double that of the first peak, and located at $L \approx 3\sigma_+$ for $\overline{\mathcal{V}} = 0$ V. The equilibrium profiles for oppositely signed wall potentials are identical, but with the cation and anion densities swapped. The differences between the corresponding pairs of curves in Figure \ref{FSW_Vscan_plots}(a,c) are solely due to the dissimilarity of the single-particle diffusivities for ionic liquid I (found in Table \ref{physproptable}). For small pores and large potentials, there is a negligible density of co-ions in the pore. In such cases, the ratio of time constants in Figure \ref{FSW_Vscan_plots}(a,c) is approximately the ratio of the diffusivities.

The plots for ionic liquid II are shown in Figure \ref{FSW_Vscan_plots}(b,d). Due to the asymmetry of the ion sizes, the equilibrium profiles for oppositely signed $\overline{\mathcal{V}}$ cannot be related as for ionic liquid I. Furthermore, the potential of zero charge is a non-zero function of $L$ (between $-0.1$ V and $0$ V for our case). To investigate the influence of the small variation in diffusivities for ionic liquid II, we computed the impedance response for $D_+$ and $D_-$ set equal to the mean of the diffusivities in Table \ref{physproptable}. When the density of one species in the pore was negligible, the response times were rescaled according to the change of the single-particle diffusivity of the dominant species. Otherwise, there was an insignificant deviation from the curves shown in Figure \ref{FSW_Vscan_plots}(b,d). In panel (b), we see smooth variation across the dashed vertical line between $\tau_{\infty}$ and the the exact single-species result, as for the BFD and fMSA results in Figure \ref{FSW_tau_plots}(b). For negative wall potentials in panel (d), the density of the smaller anions are negligible either side of $L = \sigma_+$, and so the value of $\tau_{\infty}$ is discontinuous as $\overline{N}_+$ jumps to zero. The values of $\tau_{\infty}$ to the left of the dashed line are not plotted as they correspond to a negligible quantity of ions. As observed for ionic liquid I, there are two large response time regions for the $L$-interval plotted, which shift as the base voltage varies.


\section{Discussion and conclusions\label{Conclu}}

We constructed a reduced-order limit model for the dynamics of ionic liquids in long slit pores, incorporating DFT for the equilibrium system and DDFT for the out-of-equilibrium time-evolution. We utilised this model to investigate the dependence of the impedance response on system parameters and the choice of DFT used. A semi-analytical approach was presented for the rapid evaluation of the complex impedance, with the only numerical inputs being sensitivities of the equilibrium DFT system. The derived expressions were validated with fully time-dependent numerical simulations.

The impedance response of our reduced-order model can deviate significantly from the de Levie result \eqref{deLeviemodel}, to an extent similar to that found for non-trivial pore geometries with an idealised electrolyte model \citep{cooper2017simulated}. In the high and low frequency limits, we are able to obtain effective FSW elements. We found that the inclusion of electrostatic correlations (which are vital for a physically correct description of the system \citep{voukadinova2019energetics}) drastically modified the predicted high and low frequency response times for pores with widths comparable to one or two ion diameters (see Figure \ref{FSW_tau_plots}). By varying the base voltage (at which the porescale PEIS was performed), we found that peaks in the FSW time constants shifted towards smaller pore widths as the potential difference magnitude was decreased (see Figure \ref{FSW_Vscan_plots}).

\begin{figure}
\centering
\begin{subfigure}{3in}\caption{$\overline{\mathcal{V}} \geq 0$ (I).}
\includegraphics[width=2.75in]{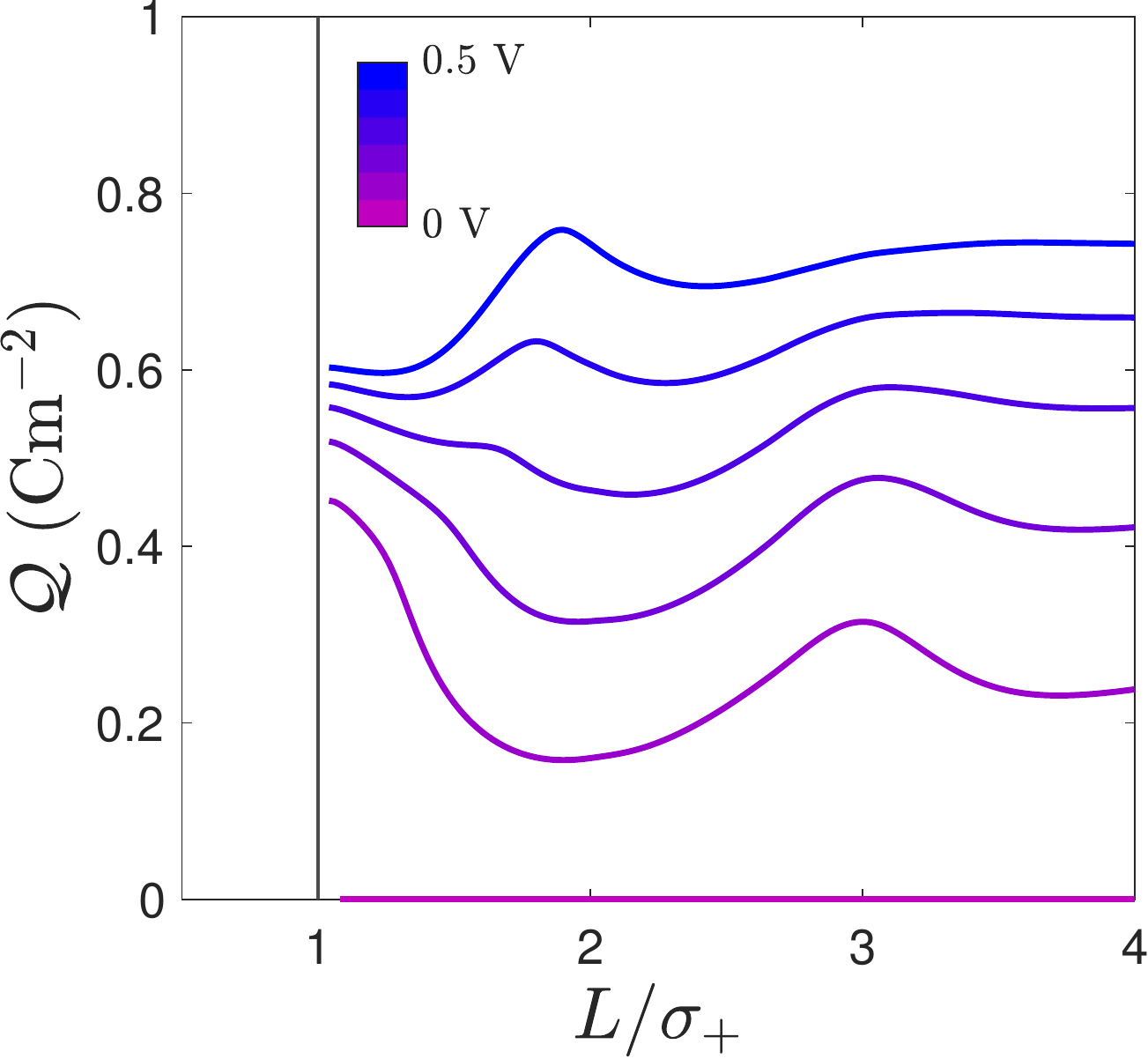}
\vspace{3mm}
\end{subfigure}
\begin{subfigure}{3in}\caption{$\overline{\mathcal{V}} \geq 0$ (II).}
\includegraphics[width=2.75in]{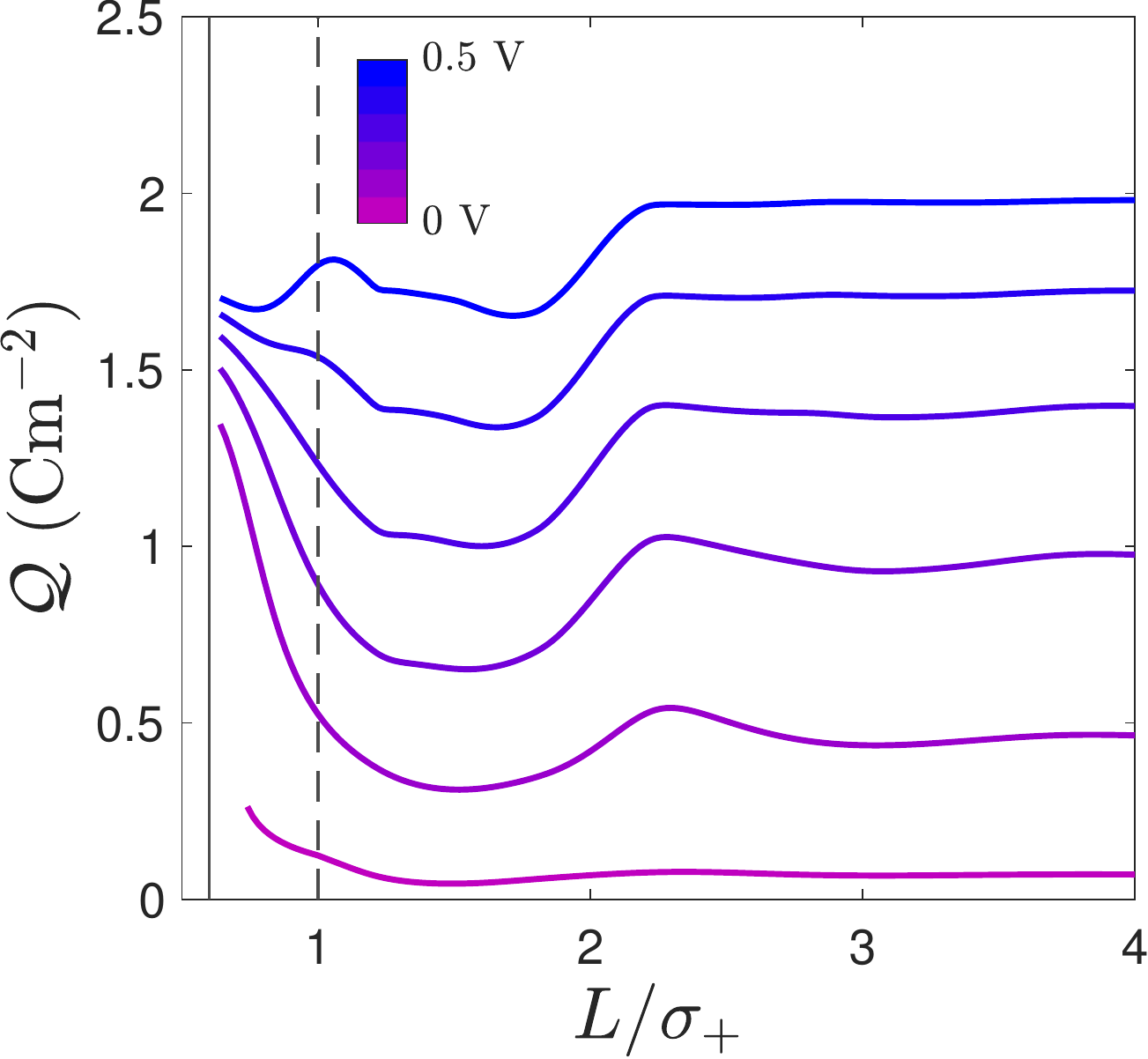}
\vspace{3mm}
\end{subfigure}
\caption{Charge stored at equilibrium (per unit of pore length and depth) against pore width for a range of $\overline{\mathcal{V}} \geq 0$ with fMSA for the DFT. The curves are analogous to those of $\tau_{\infty}$ in Figure \ref{FSW_Vscan_plots}(a,b) -- see the caption for details. Note the difference in vertical axes between the two panels.}\label{Qstored}
\end{figure}

As shown in Figure \ref{Qstored}, the charge stored in the pore also varies strongly for small widths as the underlying equilibrium changes. The oscillating dependence on $L$ becomes more dramatic for smaller values of $\epsilon_{\textrm{r}}$ (see the integral capacitance plots of \cite{jiang2011oscillation} with $\epsilon_{\textrm{r}} = 1$, for example). For supercapacitor geometries, a good distribution of pore sizes will provide an optimal balance of total charge storage and charge accessibility for a given application (which determines typical voltage and frequency ranges). For high frequency applications, the total charge stored by the individual electrode pores is relatively less important than their characteristic response times. In addition, the optimisation must respect the total pore fraction available, the fact that thinner pores take up a smaller proportion of the latter, and the shapes of pore size distribution that are feasible to manufacture.

For a positive electrode, Figure \ref{FSW_Vscan_plots}(a,b) and Figure \ref{Qstored} describe the dynamic and static charge storage properties of slit pores in the high frequency limit, respectively. For ionic liquid I, we find that peaks $\tau_{\infty}$ and $\mathcal{Q}$ coincide for $\overline{\mathcal{V}} = 0.5$ V, and so the two optimisation objectives are in conflict (improvements in total capacity result in reduced accessibility).  Slit pores with $L \approx \sigma_+$ have lower response times without a significant reduction in charge stored. For ionic liquid II, the optimisation objectives appear to be non-conflicting. Pores of width close to $\sigma_-$ provide large values of $\mathcal{Q}$ as well as fast response times, and for $L \approx 1.5 \sigma_+$, both the static and dynamic charge storage properties are relatively poor. It is clear that the presented theory (along with consideration of nonlinear responses \citep{aslyamov2020relation}) adds to the pool of information on which the optimisation of supercapacitor geometries can be based.





The model for the ionic liquid can be generalised with ease to incorporate non-constant permittivities and diffusivities; molecular dynamics and mean-field simulations have shown that confinement can cause significant variations in these quantities \citep{terrones2016enhanced,kondrat2014accelerating}. Relevant DDFT studies include the work of \cite{qing2020effects} incorporating density-dependent permittivities, and the theoretical investigation of packing fraction-dependent diffusivities by \cite{stopper2018bulk}. Inclusion of tortuosity effects is also kept for future work. Overall, the approach presented here is clearly applicable to more general particle systems or other models of ionic liquids; the asymptotic reduction of the DDFT system is an application of lubrication theory, and can be carried out for a large class of dynamical equations in long slit pores. The resulting limit model may then also be amenable to a semi-analytical impedance calculation.

We believe that modifications to the DDFT will be required to accurately model ion dynamics in long cylindrical pores. As discussed by \cite{te2020classical}, for the simple case of a single species hard-sphere fluid in single-file transport, particles can pass through each other when the time evolution is governed by the DDFT equation \eqref{DDFT3D}. In a slit pore, on the other hand, it can be thought that the ions slide past each other in the unbounded direction perpendicular to the $x$--$y$ plane. 
Extensions of \eqref{DDFT3D} may be required to simulate pore systems further from equilibrium (e.g.~nonlinear electrochemical impedance spectroscopy). One option is provided by the recently developed power functional theory \citep{schmidt2013power} which generalises standard DDFT by including non-adiabatic effects ignored by the DDFT assumption \eqref{EL0}.

Lastly, we note that the limit model presented can be used as the basis of a simplified cell-level model, in which a set of pores representing each electrode are directly coupled to the separator region (with a modification of the entrance boundary condition \eqref{entranceboundary}), akin to the simplified model presented in \cite{kondrat2013charging} which ignores the pore-size hierarchy typical of electrode geometries; this parallels the many-particle models studied by \cite{kirk2020modelling} for Li-ion batteries. The time-dependent simulation approach detailed in Appendix \ref{numericstime} was only used for validation in the present work, but for geometries more complex than a single straight pore, for which the system is likely analytically intractable, time-dependent simulations may be the only viable approach. It would be of interest to improve the resolution of the system at pore entrances and branchings which could involve a hybridisation of the limit model with a 2D DDFT solver; we have also considered hybridisation with molecular simulations as done in \cite{borg2015hybrid} for compressible nanoscale flows.



\appendix

\section{Numerical details\label{numerics}}


We first discuss the solver for the 1D DFTs required to compute equilibrium solutions and their numerical sensitivities for the semi-analytical impedance expressions, and employed as a component of the numerical scheme for time-dependent simulations. We then outline the finite-volume approach for the latter, its implementation for impedance response experiments, and provide the results of validation tests.

\subsection{Equilibrium solver\label{numericsequilibrium}}

In the present work, we perform both canonical and grand canonical equilibrium calculations (predominantly the former). In the canonical case, we are given particle numbers $N_i$ and a wall surface potential $\mathcal{V}$, from which we compute 1D equilibrium densities $\rho_i(y)$ and constant chemical potentials $\mu_i$. In the grand canonical case, the $\mu_i$ are inputs and the $N_i$ can be trivially computed by integrating the output densities. Both situations are tackled with a Picard iteration (iterative substitution) scheme with a line search based on the fixed-point problem \eqref{expform1} as described by \cite{knepley2010efficient} and \cite{roth2010fundamental}, which does not require the evaluation of $\Omega$ \eqref{grandpotential} at any stage. 
During each Picard iteration, we solve the PB equation for the electrostatic potential $\psi(y)$ arising from the substitution of (\ref{expform1},\ref{electrostaticincorporation}) into \eqref{Poisson0}. This is significantly more robust than solving the Poisson equation with iteration-lagged densities on the right hand side. For canonical computations, we augment the PB iteration scheme with the integral expression for the particle numbers \eqref{particlenumberdefn}, solving for the unknown $\mu_i$ in conjunction with the electrostatic potential.
Specifically, if $\rho_i^{(k)}$ denotes the $k^{\textrm{th}}$ Picard iteration of the density for species $i$, we first build
\begin{equation}g_i^{(k)} = \exp\left[ - \beta \left( \left. \frac{\delta F^{\textrm{HS}}}{\delta \rho_i} \right|_{\rho_i^{(k)}} + \left. \frac{\delta F^{\textrm{EC}}}{\delta \rho_i} \right|_{\rho_i^{(k)}} + {{U}}_i^{\textrm{HW}}  \right) \right].\end{equation}
Given $N_i$, we solve an augmented PB system for $\rho_i^{(k+1)}(y)$ and $\mu_i^{(k+1)}$,
\begin{equation}\rho_i^{(k+1)} = g_i^{(k)} \exp\left[ - \beta \left( {\textrm{e}} z_i \psi^{(k+1)} - \mu_i^{(k+1)} \right) \right] , \quad \epsilon_0\epsilon_{\textrm{r}} \psi^{(k+1)}_{yy} = - {\textrm{e}} \sum_i z_i \rho^{(k+1)}_i, \quad N_i = \int_{0}^{L} \rho_i^{(k+1)} \; \mathrm{d}y,\end{equation}
with the boundary conditions for the electrostatic potential \eqref{PoissonBC1}. The above system can be solved with fast gradient descent algorithms as in the unaugmented case. We found this approach to be more robust than including \eqref{particlenumberdefn} into the Euler--Lagrange equations \eqref{EL0} by using a Lagrange multiplier \citep{yu2017microstructure,lutsko2018classical}. For the BFD and fMSA theories, which take a set of constant charge-neutral ion densities as inputs, we use the bulk densities external to the pore $\rho_i^{\textrm{b}}$, even in the case when the system is not in equilibrium. 

The equilibrium solver was carefully checked against results from a number of different sources (e.g. independently written DFT code for identical systems), and thus is not the subject of the numerical validations below. Throughout, the equilibria are computed using a discretisation in $y$ of 400 grid points per nanometer, giving 200 and 120 grid points across the diameters of the [EMIm$]^+$/[TFSI$]^-$ and [BF$_4$$]^-$ ions, respectively. The tolerances of the PB solver and Picard iteration were chosen to be sufficiently small that they are not limiting the accuracy of the equilibrium solution. The surface charge density on the pore walls differed by at most $2\%$ from the fully converged result.

\begin{figure}
\centering
\begin{subfigure}{3in}\caption{DCA-PB.}
\includegraphics[width=2.75in]{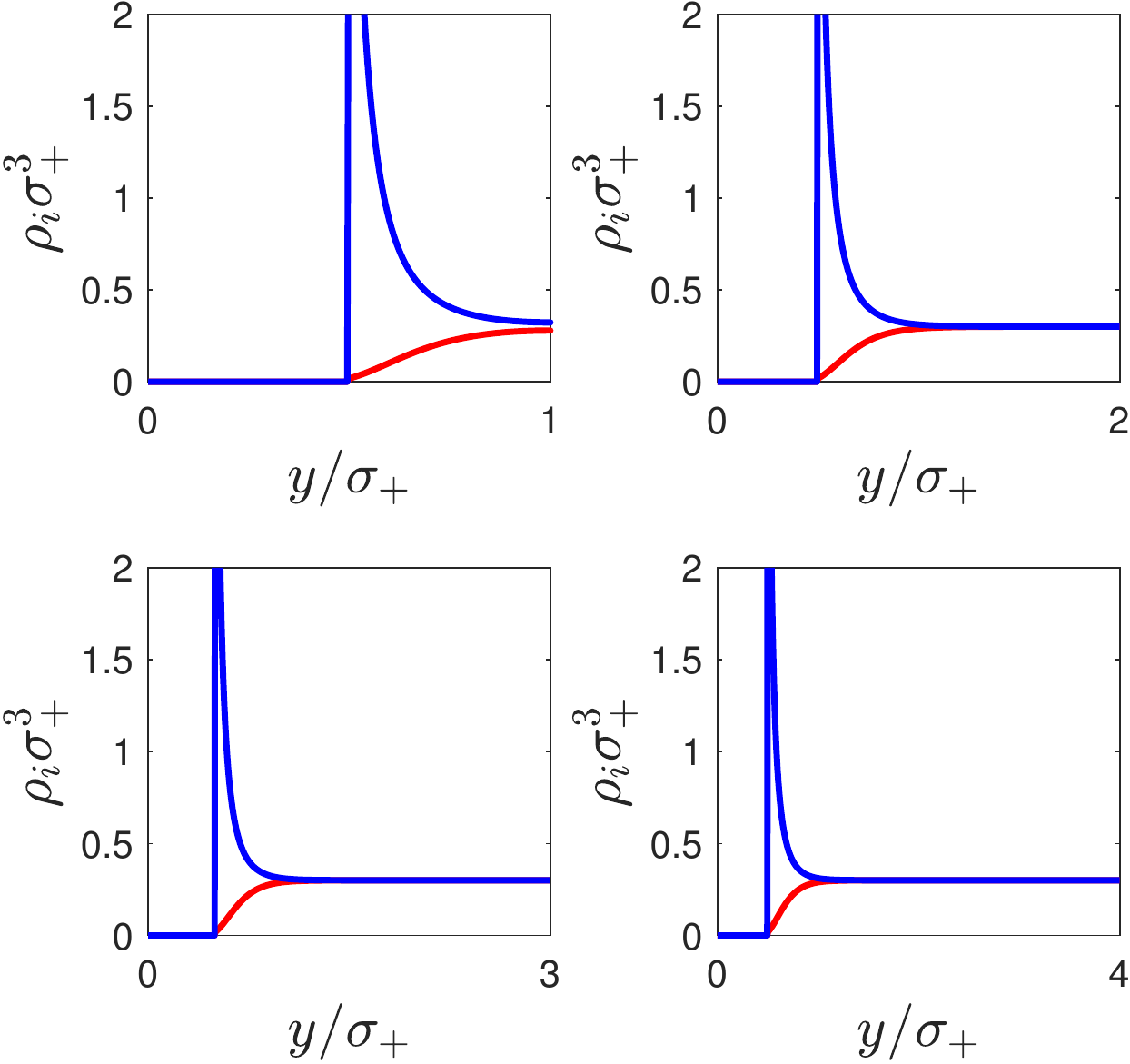}
\vspace{3mm}
\end{subfigure}
\begin{subfigure}{3in}\caption{HS-PB.}
\includegraphics[width=2.75in]{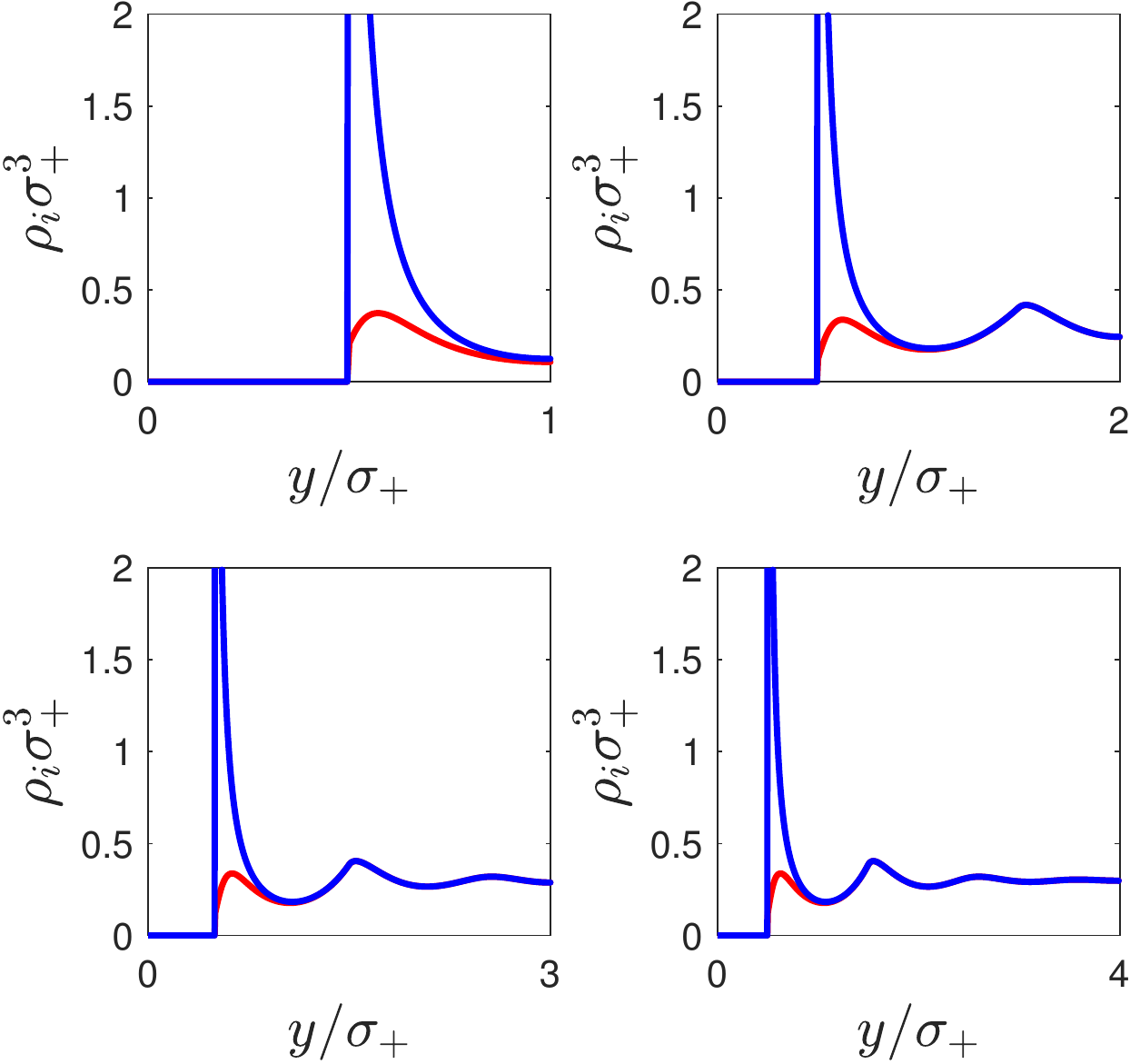}
\vspace{3mm}
\end{subfigure}
\begin{subfigure}{3in}\caption{BFD.}
\includegraphics[width=2.75in]{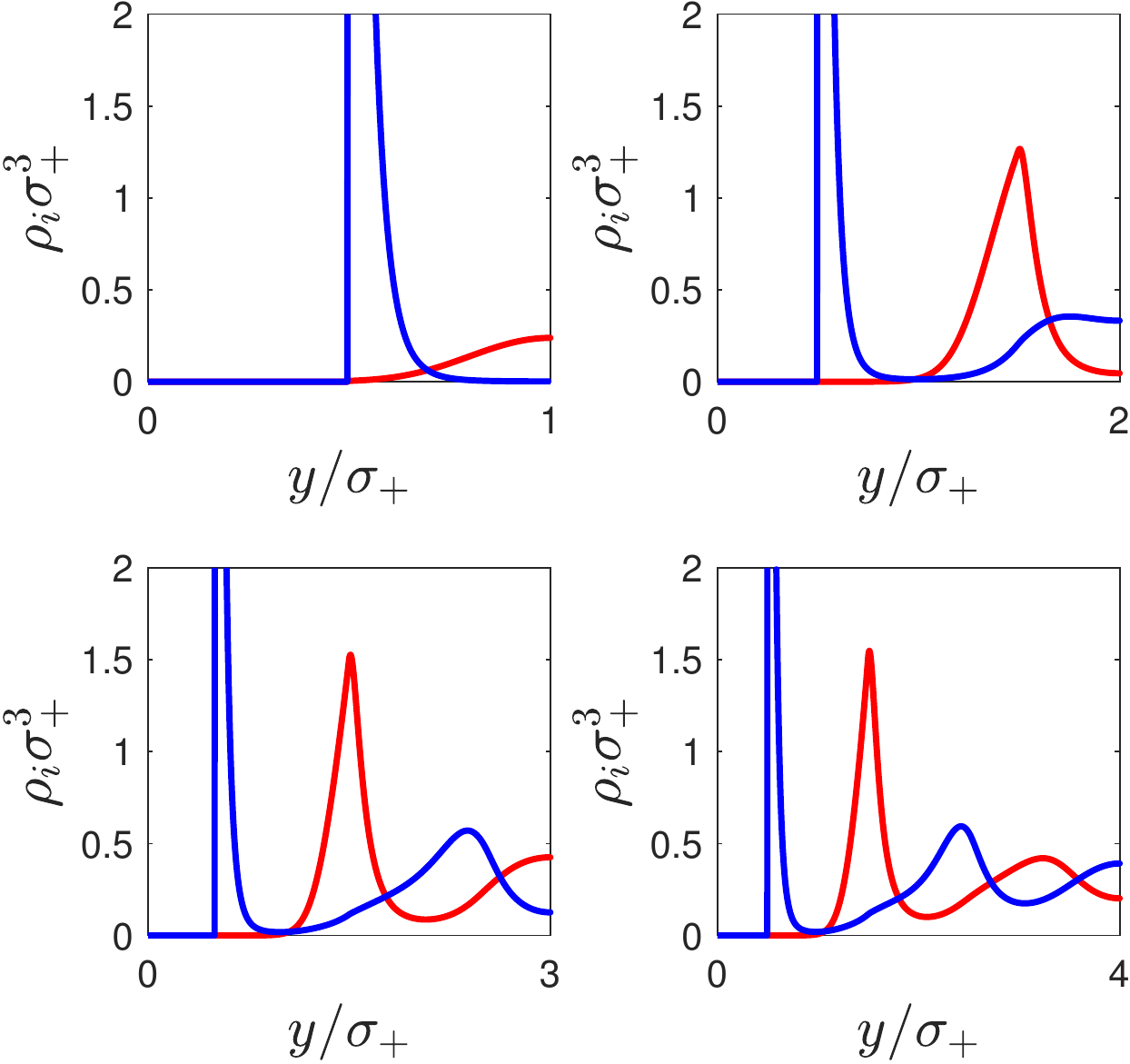}
\end{subfigure}
\begin{subfigure}{3in}\caption{fMSA.}
\includegraphics[width=2.75in]{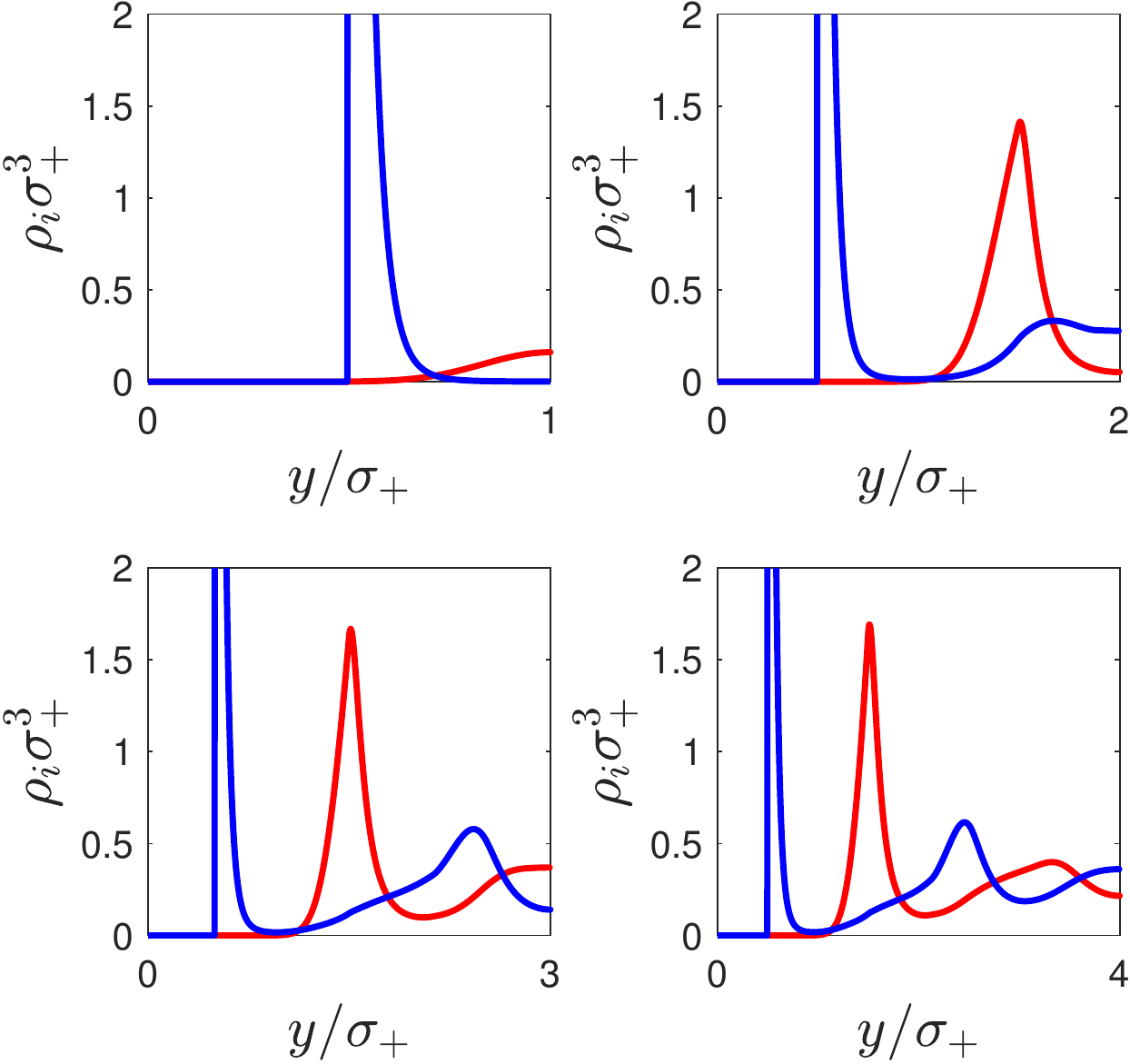}
\end{subfigure}
\caption{Equilibrium densities for the four DFTs with parameters for ionic liquid I. Each panel displays four sets of ion profiles (symmetric about the pore centerline) with $L/\sigma_+ = 2, 4, 6, 8$ (corresponding to $1,2,3,4$ nm, respectively). The potential on the pore walls is fixed at $0.5$ V. The cation/anion densities are plotted in red/blue.}\label{paramsetIequilibria}
\end{figure}

Equilibrium density profiles for ionic liquid I (see Table \ref{physproptable}) are presented in Figure \ref{paramsetIequilibria} for the four DFTs considered in this work. The left half pore is plotted as the profiles are symmetric about the pore centreline. For these grand canonical calculations, we have fixed the chemical potential equal to the bulk value (determined by $\rho_i^{\textrm{b}}\sigma_+^3 = 0.3$), and set the pore surface potential to $0.5$ V. The ion layering induced by the addition of hard-sphere terms is clear from a comparison of panels (a) and (b), while the addition of electrostatic correlations introduced in panels (c,d) disrupts the local charge neutrality seen in (a,b). There is good agreement between the two implementations of electrostatic correlations, with a slight difference in amplitude of the density profiles shown in panels (c) and (d).

\subsection{Time-dependent simulations of the limit model\label{numericstime}}

For the following, we move to a non-dimensional system by rescaling with
\begin{equation}\label{DDFTrescalings}(x,y) = \sigma_{+} (x^*, y^*), \qquad \rho_i = \sigma_{+}^{-3} \rho_i^*, \qquad  t = \frac{\sigma_+^2}{D_+} t^*,  \qquad \mu_i = \beta^{-1} \mu_i^*,\end{equation}
and dropping the stars. Here, and for our other numerical calculations, we have used the cation diameter $\sigma_{+}$, the cation diffusion time $\sigma_+^2/D_+$, and the inverse thermodynamic temperature $\beta^{-1}$ as our reference length, time, and energy scales, respectively. This particular rescaling is chosen as it is fixed for a given ionic liquid and temperature (independent of input frequency). The limit model \eqref{weakform1} becomes
\begin{equation} \label{weakform2}\frac{\partial N_i}{\partial t}  =   \mathcal{D}_i \frac{\partial}{\partial x} \left( N_i  \frac{\partial \mu_i}{\partial x} \right),
\end{equation}
where $\mathcal{D}_i = D_i/D_+$ are non-dimensional diffusion coefficients.

We discretise the slit pore in the length variable $x$ by sub-dividing into cells (finite volumes), where the end points of cell $j$ are denoted $x_{j-1/2}$ and $x_{j+1/2}$, with midpoint $x_j$. 
Starting with particle numbers in each cell at time $t$, $N_i^j(t)$, the 1D densities $\rho_i^j(y,t)$ and chemical potentials $\mu_i^j(t)$ are calculated as outlined above. The computational complexity of this step is mitigated significantly by parallelisation, and the fact that the equilibrations are initialised with the densities from the previous time-step (the profiles $\rho_i^j$ serve no further purpose in the following).

To evolve the particle numbers in time according to \eqref{weakform2}, we employ the mass-conserving finite volume scheme outlined by \cite{carrillo2015finite} which has recently been employed for the numerical simulation of a DDFT system by \cite{russo2020finite}. Multiplying \eqref{weakform2} by an indicator function for cell $j$ (defined by $\zeta^j(x) = 1$ for $x \in [x_{j-1/2}, x_{j+1/2}]$ and zero otherwise) and integrating yields an exact analytical expression for the rate of change of the total particle numbers of each ion species in cell $j$ in terms of continuous fluxes,
\begin{equation} \label{finitevolumebalance}\frac{\mathrm{d}}{\mathrm{d}t} \int_{x_{j-1/2}}^{x_{j+1/2}} N_i  \;   \mathrm{d}x  
= \mathcal{D}_i \left[ N_i  \frac{\partial \mu_i}{\partial x} \right]_{x_{j-1/2}}^{x_{j+1/2}} .
\end{equation}
Substituting $N_i = N_i^j(t)$ in the left hand side of \eqref{finitevolumebalance} and defining $\Delta x_j = x_{j+1/2} - x_{j-1/2}$, we approximate this exact integral balance with
\begin{equation}\label{finitevolumebalanceapprox} \frac{\mathrm{d}}{\mathrm{d}t} N_i^j = \frac{f_i^{j+1/2} - f_i^{j-1/2}}{\Delta x_j},
\end{equation}
where we have defined the numerical inter-cell fluxes
\begin{equation}\label{fluxexpression}f_i^{j\pm1/2} \approx \mathcal{D}_i \left. \left( N_i  \frac{\partial \mu_i}{\partial x} \right) \right|_{x_{j\pm1/2}}.
\end{equation}

To calculate the numerical fluxes, we utilise positivity preserving piecewise linear reconstructions of the particle numbers. Fixing time $t$ (and dropping its dependence from our notation), we define
\begin{equation}
\tilde{N}_i^j(x) = N_i^j + (x-x_j) {\eta}_i^j,
\end{equation}
where ${\eta}_i^j$ is the slope of the reconstruction. 
We employ the minmod limiter,
\begin{equation}\label{piecewiselinear}
 {\eta}_i^j =  \operatorname{minmod}\left(\frac{N_i^{j+1} - N_i^{j} }{ x_{j+1} - x_{j}}, \frac{N_i^j - N_i^{j-1} }{ x_{j} - x_{j-1}} \right),
\end{equation}
where the $\operatorname{minmod}$ operator is defined, as in \cite{carrillo2015finite}, by
\begin{equation}
\operatorname{minmod}( y_1, y_2, \ldots ) :=  \begin{cases} 
\min\{ y_1, y_2, \ldots \} & \textrm{if } y_k > 0 \; \forall k, \\ 
\max\{ y_1, y_2, \ldots \} & \textrm{if } y_k < 0 \; \forall k, \\ 
0 & \textrm{otherwise.}
\end{cases}
\end{equation}
In simple terms, the slope \eqref{piecewiselinear} is set equal to zero at a turning point or otherwise chosen to be the one-sided finite difference with the smallest absolute value (minimum modulus). It follows that the reconstructed $\tilde{N}_i^j(x)$ are non-negative within cell $j$. \cite{carrillo2015finite} and \cite{russo2020finite} utilise a generalised minmod limiter involving both one-sided and central finite differences with a tuneable numerical viscosity. Using such reconstructions, we can approximate $N_i$ at the cell boundaries $x_{j\pm1/2}$ required to compute the fluxes \eqref{fluxexpression}. Following the notations in \cite{carrillo2015finite}, we define the reconstructed particle numbers at the East and West boundaries of a cell,
\begin{equation}N_i^{j,\textrm{W}} = \tilde{N}_i^j(x_{j - 1/2}) = N_i^j - \frac{\Delta x_j}{2}{\eta}_i^j ,\qquad N_i^{j,\textrm{E}} = \tilde{N}_i^j(x_{j +1/2}) = N_i^j + \frac{\Delta x_j}{2} {\eta}_i^j.
\end{equation}
For the chemical potential component of \eqref{fluxexpression} we take a simple finite difference, and define the numerical velocities
\begin{equation} \label{numericalvelocities}
- u_i^{j-1/2}  = \mathcal{D}_i \left( \frac{\mu_i^j - \mu_i^{j-1}}{x_j - x_{j-1}}\right).
\end{equation}
We use an upwind scheme in space, choosing the flux 
\begin{equation} \label{numericalfluxdefn1}
-  f_i^{j-1/2}   =  u_i^{j-1/2} \begin{cases} {\displaystyle N_i^{j-1,\textrm{E}}  } & \textrm{if } u_i^{j-1/2} \geq 0, \\
 \vspace{-3mm}\\
{\displaystyle   N_i^{j,\textrm{W}}  }& \textrm{if } u_i^{j-1/2} < 0, \end{cases}
 \end{equation}
which can be written equivalently as
\begin{equation}
- f_i^{j-1/2}   =  \max\{u_i^{j-1/2},0\} N_i^{j-1,\textrm{E}}  + \min\{u_i^{j-1/2},0\} N_i^{j,\textrm{W}} .
\end{equation}

The positivity of the particle numbers can be ensured if strongly stability preserving (SSP) schemes are used for the time discretisation, subject to the appropriate Courant--Friedrichs--Lewy (CFL) condition. We use the forward Euler scheme with the CFL condition
 \begin{equation}\label{CFLpositivity} \Delta t \cdot \max_j\left\{ \frac{u_i^{j+1/2}}{ \Delta x_j} , - \frac{u_i^{j-1/2}}{ \Delta x_j}\right\} \leq \frac{1}{2}.
\end{equation} 
This result is valid for any choice of positivity preserving piecewise linear reconstruction, and the proof relies on the useful property that $N_i^j = (N_i^{j,\textrm{E}} + N_i^{j,\textrm{W}})/2$. The time-step is chosen adaptively to satisfy the dynamic constraint \eqref{CFLpositivity}, however this CFL condition is not sufficient to ensure stability of the numerical scheme. We performed validation tests with higher-order multi-step SSP Runge--Kutta schemes, which are too computationally expensive to be used instead of the Euler scheme due to the increased number of DFT solves without a significant easing of time-step requirements.

\subsubsection{Implementation for frequency response simulations}

Our implementation of the above numerical scheme is tuned for frequency response simulations; a different set-up may be more appropriate for other experiment types. For a given frequency, the perturbation to the particle numbers reaches a distance into the pore characterised by the penetration depth $h_{\textrm{p}} = \sqrt{\max \{ D_i \} /\omega}$. At sufficiently low frequencies for which $h_{\textrm{p}} \approx {H}$, the no-flux boundary at the pore cap begins to influence the dynamics and the resulting impedance behaviour \citep{keiser1976abschatzung,cooper2017simulated}. In order to attain a numerically efficient discretisation of the pore, the cell widths, $\Delta x_j$, should be chosen according to $h_{\textrm{p}}$. For this, we use a convex stretching function which maps a set of equidistant points in the interval $[0,1]$ to a distribution of cell midpoints in $[0,{H}]$, providing three levels of spatial resolution in the pore length; the highest resolution is attained in a layer of width $4 h_{\textrm{p}}$ adjacent to the pore entrance, good resolution in the region $[4h_{\textrm{p}},75h_{\textrm{p}}]$, and sparsely distributed cell midpoints beyond this (as the densities remain relatively close to equilibrium in this region). As $h_{\textrm{p}}$ increases, the three grades of resolution transition to two, and eventually the spatial discretisation becomes uniform. In order to obtain convergent numerical simulations, we need a more strict time-step constraint than the CFL condition \eqref{CFLpositivity}. For a fixed spatial discretisation, the time-step required for numerical stability depends on the frequency imposed. At high frequencies, the number of time-steps per period must be sufficiently large, whereas for low frequencies, a frequency-independent time-step is needed; explicitly, in addition to \eqref{CFLpositivity}, we use a constraint of the form
\begin{equation} \Delta t \leq \min \{ \Delta t_1, 1/ f \nu \},\end{equation}
where $\Delta t_1$ denotes the frequency-independent upper bound on the time-step and $\nu$ is the desired minimum number of time-steps per period.

\begin{figure}
\centering
\begin{subfigure}{2.8in}\caption{$\mathcal{V}(t) = \overline{\mathcal{V}}$, $\mathcal{V}'(t)>0$}
\includegraphics[width=2.55in]{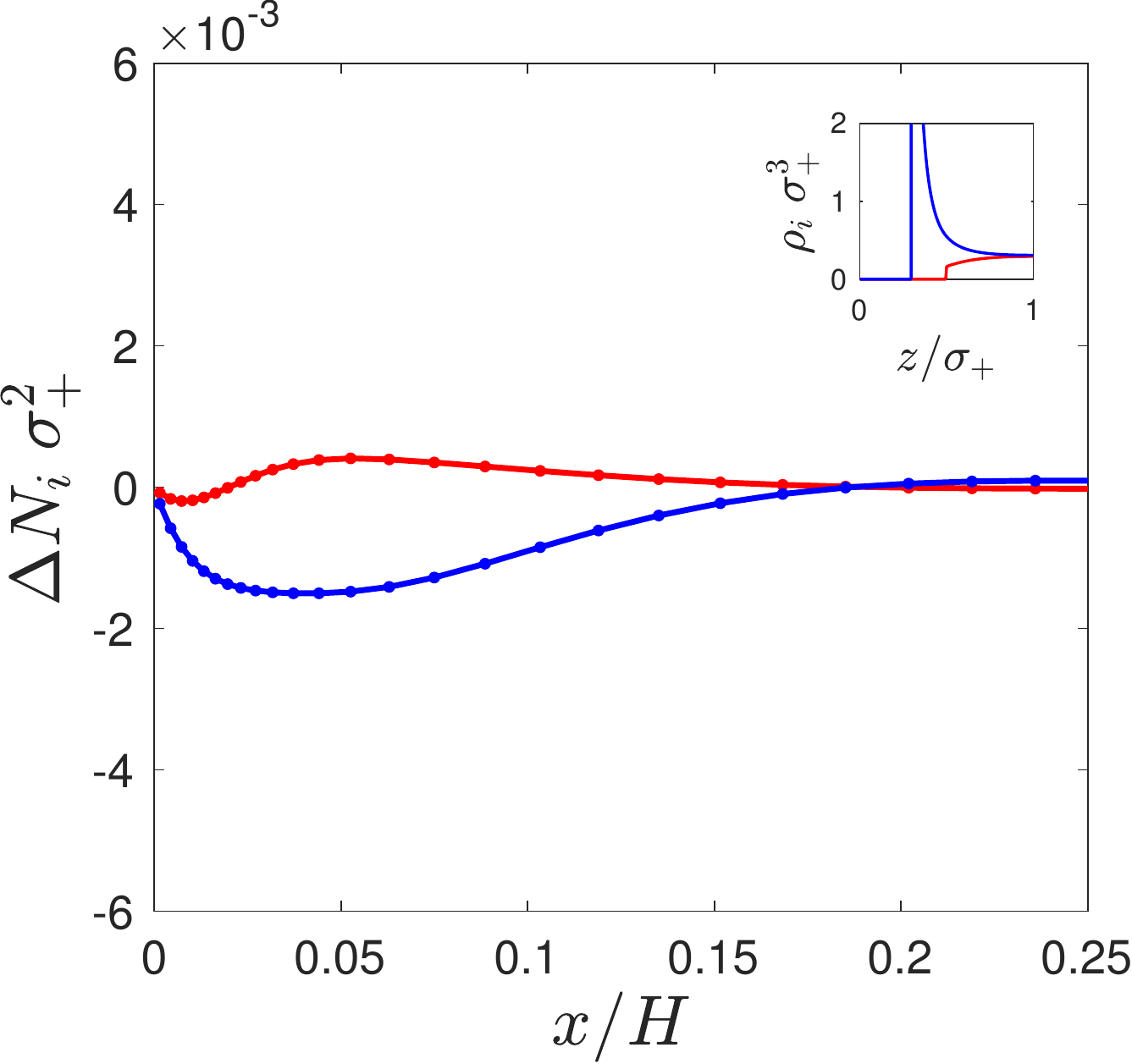}
\end{subfigure}
\begin{subfigure}{2.8in}\caption{$\mathcal{V}(t) = \overline{\mathcal{V}} + {\max_t|\Delta \mathcal{V}(t)|}$}
\includegraphics[width=2.55in]{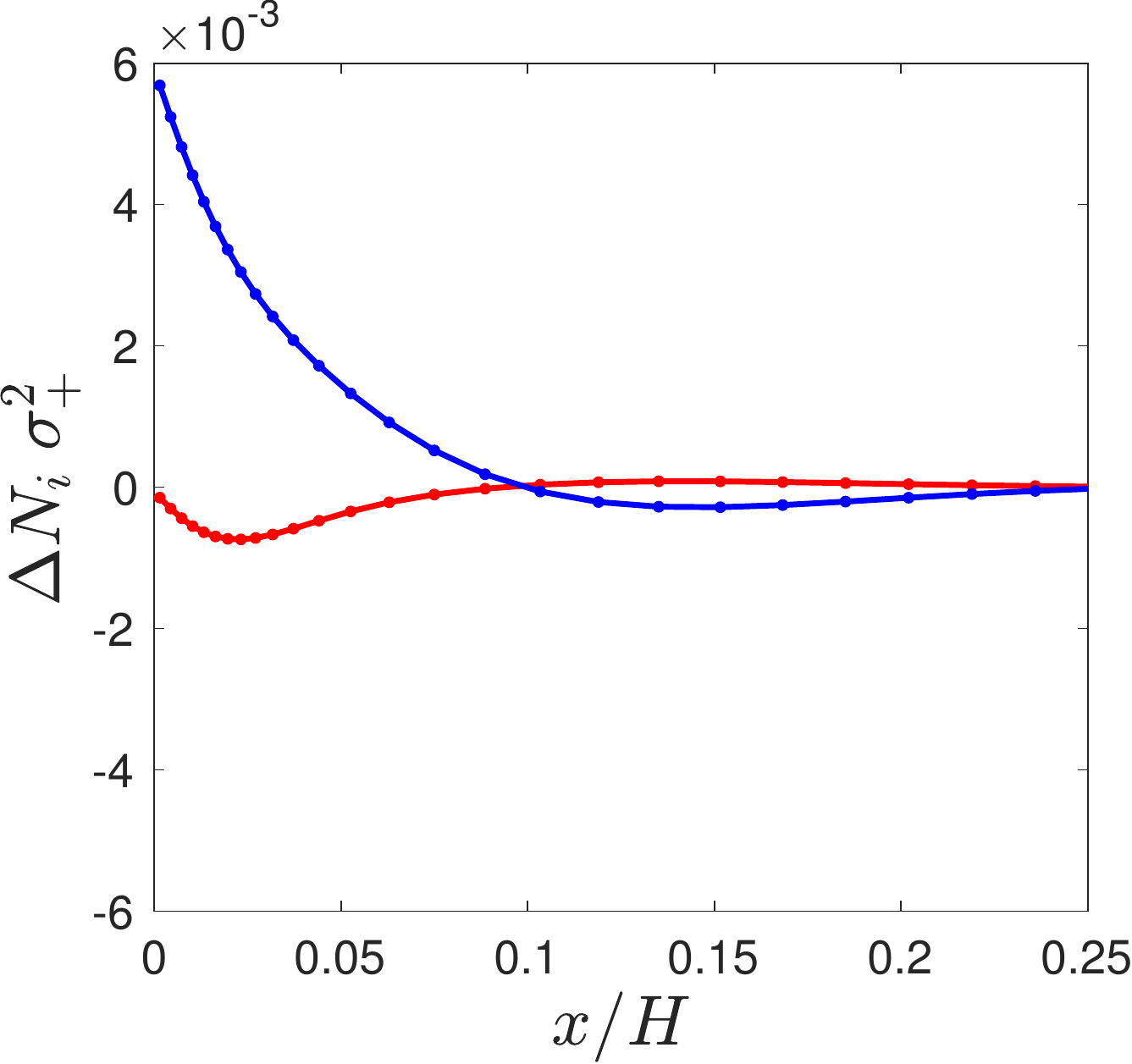}
\end{subfigure}
\vskip 4mm
\begin{subfigure}{2.8in}\caption{$\mathcal{V}(t) = \overline{\mathcal{V}}$, $\mathcal{V}'(t)<0$}
\includegraphics[width=2.55in]{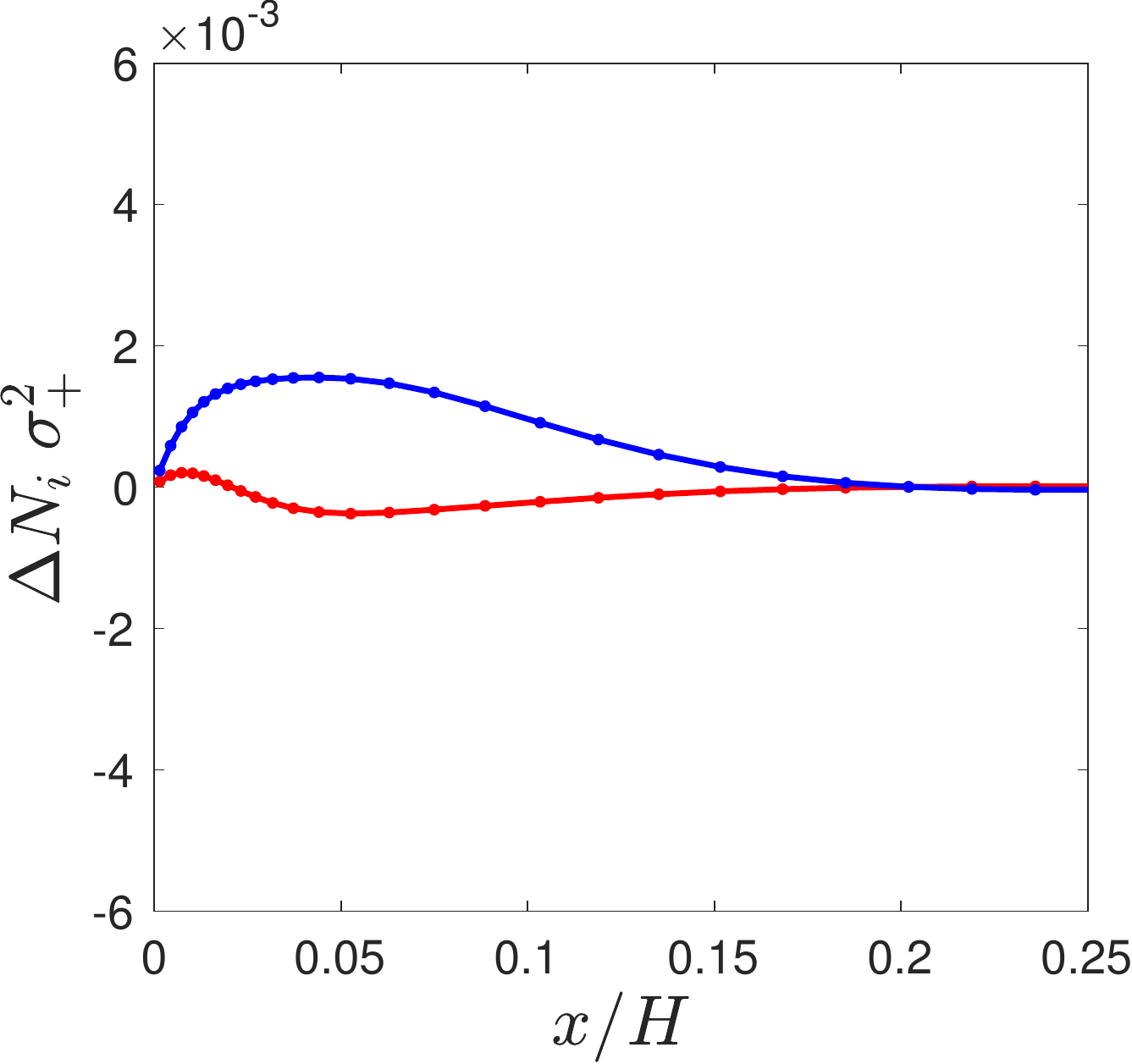}
\end{subfigure}
\begin{subfigure}{2.8in}\caption{$\mathcal{V}(t) = \overline{\mathcal{V}} - {\max_t|\Delta \mathcal{V}(t)|}$}
\includegraphics[width=2.55in]{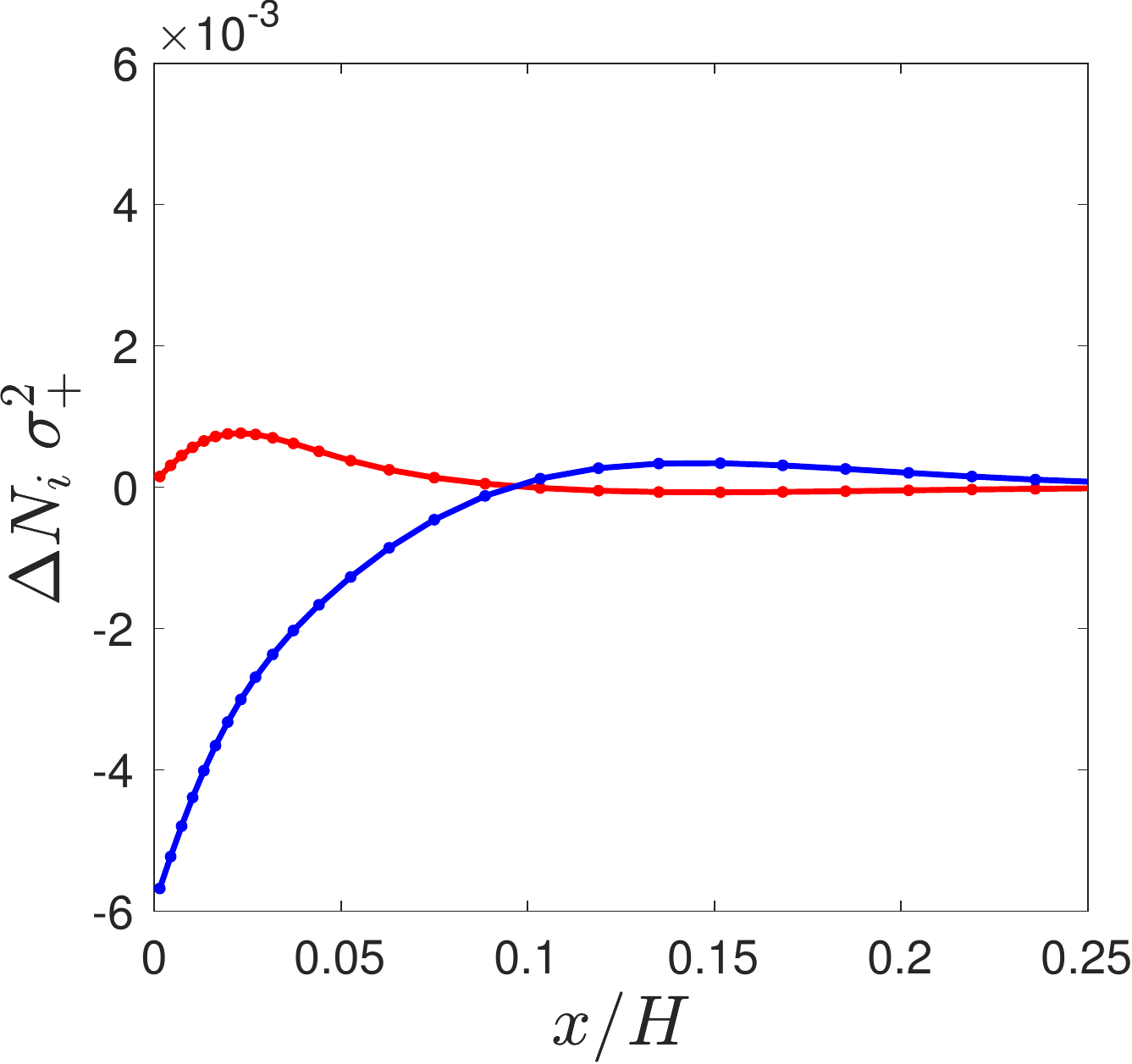}
\end{subfigure}
\caption{Perturbation particle numbers at different stages of a frequency response simulation with $f = 10$ Hz. The slit pore considered has width ${L} = 1$ nm and length ${H} = 100$ $\mu$m, and the applied wall potential is defined by $\overline{\mathcal{V}} = 0.5$ V, ${\max_t|\Delta \mathcal{V}(t)|} = 2.5$ mV. We use the parameters for ionic liquid II from Table \ref{physproptable} and the DCA-PB model, with the unperturbed equilibrium profile inset in panel (a). The cation/anion particle numbers are plotted in red/blue.}\label{DCAPBplots}
\end{figure}

In Figure \ref{DCAPBplots} we present snapshots of the particle number perturbations in a slit pore undergoing a frequency response simulation. The four panels correspond to the two zeros and two peaks of the voltage perturbation in a single period. The density profiles inset in panel (a) show that the pore holds more anions than cations, with $\overline{N}_+ \sigma_+^2 = 0.260$ and $\overline{N}_- \sigma_+^2 = 1.34$, and all panels show that the anion deviation $\Delta N_{-} \sigma_+^2$ (plotted in blue) is generally of larger amplitude than the cation deviation. The different levels of numerical resolution can be seen from the clustering of the cell midpoints, indicated by the points on the curves.

\subsubsection{Numerical validation}

\begin{table}
\begin{tabular}{cc|c|c|c|c|l}
\cline{3-5}
& & \multicolumn{3}{ c| }{\# of time-steps per period} \\ \cline{3-5}
& & 250 & 1000 & 4000 \\ \cline{1-5}
\multicolumn{1}{ |c  }{\multirow{4}{*}{\# of cells} } &
\multicolumn{1}{ |c| }{8} & $0.1996 - 0.1705\textrm{i}$ & $0.1988 - 0.1716\textrm{i}$ & $0.1985 - 0.1719\textrm{i}$  \\ \cline{2-5}
\multicolumn{1}{ |c  }{}                        &
\multicolumn{1}{ |c| }{16} & $0.1775 - 0.1723\textrm{i}$ & $0.1766 - 0.1731\textrm{i}$ & $0.1764 - 0.1733\textrm{i}$    \\ \cline{2-5}
\multicolumn{1}{ |c  }{}                        &
\multicolumn{1}{ |c| }{32} & -- & $0.1710 - 0.1702\textrm{i}$ & $0.1708 - 0.1704\textrm{i}$   \\ \cline{2-5}
\multicolumn{1}{ |c  }{}                        &
\multicolumn{1}{ |c| }{64} & -- & -- & $0.1695 - 0.1697\textrm{i}$  \\ \cline{1-5}
\end{tabular}
\caption{Numerical convergence of the complex impedance ($\Omega$\,m$^2$) for the same parameters as in Figure \ref{DCAPBplots} -- see the caption for details. Dashes denote non-convergence. The semi-analytical result is approximately $0.1695 - 0.1695\textrm{i}$.}\label{convergetable}
\end{table}

In Table \ref{convergetable}, we provide the calculated complex impedance (as briefly discussed in Section \ref{timedepsection}) for the same case as in Figure \ref{DCAPBplots} with different spatial and temporal discretisations. The time-step is proportional to the reciprocal of the number of time-steps per period (columns), and the non-uniform spatial increments are proportional to the reciprocal of the number of cells (rows). As expected for a system with similarities to the standard diffusion equation, increasing the spatial resolution by a factor of $n$ (down the rows of Table \ref{convergetable}) requires a decrease in the time-step by a factor of $n^2$ (across the columns) for the scheme to converge numerically. As long as the time-step is sufficiently small for numerical stability, decreasing it further does not result in a significant change to the computed complex impedance. With 32 cells and 1000 time-steps per period, the complex impedance deviates by less than 1\% from the semi-analytical result, $Z \approx 0.1695 - 0.1695\textrm{i}$. For the results of the time-dependent simulations presented in the main text, we typically discretised the pore with 32 or 64 cells, depending on the numerical complexity of the DFT to be solved in each cell. For very low frequencies only 16 finite volumes are used as the particle numbers along the pore are close to uniform and otherwise the simulation times become excessively long.

\bibliographystyle{plainnat}
\bibliography{references}

\begin{thebibliography}{77}
\providecommand{\natexlab}[1]{#1}
\providecommand{\url}[1]{\texttt{#1}}
\expandafter\ifx\csname urlstyle\endcsname\relax
  \providecommand{\doi}[1]{doi: #1}\else
  \providecommand{\doi}{doi: \begingroup \urlstyle{rm}\Url}\fi

\bibitem[Aslyamov et~al.(2020)Aslyamov, Sinkov, and
  Akhatov]{aslyamov2020relation}
Timur Aslyamov, Konstantin Sinkov, and Iskander Akhatov.
\newblock Relation between charging times and storage properties of nanoporous
  supercapacitors.
\newblock \emph{arXiv preprint arXiv:2011.04575}, 2020.

\bibitem[Babel et~al.(2018)Babel, Eikerling, and L{\"o}wen]{babel2018impedance}
Sonja Babel, Michael Eikerling, and Hartmut L{\"o}wen.
\newblock Impedance resonance in narrow confinement.
\newblock \emph{The Journal of Physical Chemistry C}, 122\penalty0
  (38):\penalty0 21724--21734, 2018.

\bibitem[Bazant et~al.(2011)Bazant, Storey, and Kornyshev]{bazant2011double}
Martin~Z Bazant, Brian~D Storey, and Alexei~A Kornyshev.
\newblock Double layer in ionic liquids: {O}verscreening versus crowding.
\newblock \emph{Physical review letters}, 106\penalty0 (4):\penalty0 046102,
  2011.

\bibitem[Blum(1975)]{blum1975mean}
L~Blum.
\newblock Mean spherical model for asymmetric electrolytes: I. method of
  solution.
\newblock \emph{Molecular Physics}, 30\penalty0 (5):\penalty0 1529--1535, 1975.

\bibitem[Borg et~al.(2015)Borg, Lockerby, and Reese]{borg2015hybrid}
Matthew~K Borg, Duncan~A Lockerby, and Jason~M Reese.
\newblock A hybrid molecular--continuum method for unsteady compressible
  multiscale flows.
\newblock \emph{Journal of Fluid Mechanics}, 768:\penalty0 388--414, 2015.

\bibitem[Borukhov et~al.(1997)Borukhov, Andelman, and
  Orland]{borukhov1997steric}
Itamar Borukhov, David Andelman, and Henri Orland.
\newblock Steric effects in electrolytes: A modified {P}oisson--{B}oltzmann
  equation.
\newblock \emph{Physical review letters}, 79\penalty0 (3):\penalty0 435, 1997.

\bibitem[Carrillo et~al.(2015)Carrillo, Chertock, and
  Huang]{carrillo2015finite}
Jos{\'e}~A Carrillo, Alina Chertock, and Yanghong Huang.
\newblock A finite-volume method for nonlinear nonlocal equations with a
  gradient flow structure.
\newblock \emph{Communications in Computational Physics}, 17\penalty0
  (1):\penalty0 233--258, 2015.

\bibitem[Chacon et~al.(2014)Chacon, del Castillo-Negrete, and
  Hauck]{chacon2014asymptotic}
Luis Chacon, Diego del Castillo-Negrete, and Cory~D Hauck.
\newblock An asymptotic-preserving semi-{L}agrangian algorithm for the
  time-dependent anisotropic heat transport equation.
\newblock \emph{Journal of Computational Physics}, 272:\penalty0 719--746,
  2014.

\bibitem[Cooper et~al.(2017)Cooper, Bertei, Finegan, and
  Brandon]{cooper2017simulated}
Samuel~J Cooper, Antonio Bertei, Donal~P Finegan, and Nigel~P Brandon.
\newblock Simulated impedance of diffusion in porous media.
\newblock \emph{Electrochimica Acta}, 251:\penalty0 681--689, 2017.

\bibitem[de~Las~Heras and Schmidt(2014)]{de2014full}
Daniel de~Las~Heras and Matthias Schmidt.
\newblock Full canonical information from grand-potential density-functional
  theory.
\newblock \emph{Physical review letters}, 113\penalty0 (23):\penalty0 238304,
  2014.

\bibitem[De~Levie(1964)]{de1964porous}
R~De~Levie.
\newblock On porous electrodes in electrolyte solutions--{IV}.
\newblock \emph{Electrochimica acta}, 9\penalty0 (9):\penalty0 1231--1245,
  1964.

\bibitem[De~Levie(1963)]{de1963porous}
Robert De~Levie.
\newblock On porous electrodes in electrolyte solution--{I},{II},{III}.
\newblock \emph{Electrochimica acta}, 8\penalty0 (10):\penalty0 751--780, 1963.

\bibitem[De~Levie(1967)]{de1967electrochemical}
Robert De~Levie.
\newblock Electrochemical response of porous and rough electrodes.
\newblock \emph{Advances in electrochemistry and electrochemical engineering},
  6:\penalty0 329--397, 1967.

\bibitem[Dieterich et~al.(1990)Dieterich, Frisch, and
  Majhofer]{dieterich1990nonlinear}
W~Dieterich, HL~Frisch, and A~Majhofer.
\newblock Nonlinear diffusion and density functional theory.
\newblock \emph{Zeitschrift f{\"u}r Physik B Condensed Matter}, 78\penalty0
  (2):\penalty0 317--323, 1990.

\bibitem[Evans(1979)]{evans1979nature}
Robert Evans.
\newblock The nature of the liquid-vapour interface and other topics in the
  statistical mechanics of non-uniform, classical fluids.
\newblock \emph{Advances in Physics}, 28\penalty0 (2):\penalty0 143--200, 1979.

\bibitem[Evans(1992)]{evans1992density}
Robert Evans.
\newblock Density functionals in the theory of nonuniform fluids.
\newblock \emph{Fundamentals of inhomogeneous fluids}, 1:\penalty0 85--176,
  1992.

\bibitem[Gillespie(2015)]{gillespie2015review}
Dirk Gillespie.
\newblock A review of steric interactions of ions: {W}hy some theories succeed
  and others fail to account for ion size.
\newblock \emph{Microfluidics and Nanofluidics}, 18\penalty0 (5-6):\penalty0
  717--738, 2015.

\bibitem[Gillespie et~al.(2002)Gillespie, Nonner, and
  Eisenberg]{gillespie2002coupling}
Dirk Gillespie, Wolfgang Nonner, and Robert~S Eisenberg.
\newblock Coupling {P}oisson--{N}ernst--{P}lanck and density functional theory
  to calculate ion flux.
\newblock \emph{Journal of Physics: Condensed Matter}, 14\penalty0
  (46):\penalty0 12129, 2002.

\bibitem[Gillespie et~al.(2003)Gillespie, Nonner, and
  Eisenberg]{gillespie2003density}
Dirk Gillespie, Wolfgang Nonner, and Robert~S Eisenberg.
\newblock Density functional theory of charged, hard-sphere fluids.
\newblock \emph{Physical Review E}, 68\penalty0 (3):\penalty0 031503, 2003.

\bibitem[Hansen-Goos and Roth(2006)]{hansen2006density}
Hendrik Hansen-Goos and Roland Roth.
\newblock Density functional theory for hard-sphere mixtures: the {W}hite
  {B}ear version mark {II}.
\newblock \emph{Journal of Physics: Condensed Matter}, 18\penalty0
  (37):\penalty0 8413, 2006.

\bibitem[Hoffmann and Gillespie(2013)]{hoffmann2013ion}
Jordan Hoffmann and Dirk Gillespie.
\newblock Ion correlations in nanofluidic channels: {E}ffects of ion size,
  valence, and concentration on voltage-and pressure-driven currents.
\newblock \emph{Langmuir}, 29\penalty0 (4):\penalty0 1303--1317, 2013.

\bibitem[Huang et~al.(2020)Huang, Gao, Luo, Wang, Li, Chen, and
  Zhang]{huang2020review}
Jun Huang, Yu~Gao, Jin Luo, Shangshang Wang, Chenkun Li, Shengli Chen, and
  Jianbo Zhang.
\newblock Review -- {I}mpedance {R}esponse of {P}orous {E}lectrodes:
  {T}heoretical {F}ramework, {P}hysical {M}odels and {A}pplications.
\newblock \emph{Journal of the Electrochemical Society}, 2020.

\bibitem[Huang et~al.(2011)Huang, Jiang, Sasisanker, Driver, and
  Weing{\"a}rtner]{huang2011static}
Mian-Mian Huang, Yanping Jiang, Padmanabhan Sasisanker, Gordon~W Driver, and
  Hermann Weing{\"a}rtner.
\newblock Static relative dielectric permittivities of ionic liquids at 25 {C}.
\newblock \emph{Journal of Chemical \& Engineering Data}, 56\penalty0
  (4):\penalty0 1494--1499, 2011.

\bibitem[Hunger et~al.(2009)Hunger, Stoppa, Schr{\"o}dle, Hefter, and
  Buchner]{hunger2009temperature}
Johannes Hunger, Alexander Stoppa, Simon Schr{\"o}dle, Glenn Hefter, and
  Richard Buchner.
\newblock Temperature dependence of the dielectric properties and dynamics of
  ionic liquids.
\newblock \emph{ChemPhysChem}, 10\penalty0 (4):\penalty0 723--733, 2009.

\bibitem[Jiang et~al.(2011{\natexlab{a}})Jiang, Jin, and
  Wu]{jiang2011oscillation}
De-en Jiang, Zhehui Jin, and Jianzhong Wu.
\newblock Oscillation of capacitance inside nanopores.
\newblock \emph{Nano letters}, 11\penalty0 (12):\penalty0 5373--5377,
  2011{\natexlab{a}}.

\bibitem[Jiang et~al.(2011{\natexlab{b}})Jiang, Meng, and Wu]{jiang2011density}
De-en Jiang, Dong Meng, and Jianzhong Wu.
\newblock Density functional theory for differential capacitance of planar
  electric double layers in ionic liquids.
\newblock \emph{Chemical Physics Letters}, 504\penalty0 (4-6):\penalty0
  153--158, 2011{\natexlab{b}}.

\bibitem[Jiang et~al.(2012)Jiang, Jin, Henderson, and Wu]{jiang2012solvent}
De-en Jiang, Zhehui Jin, Douglas Henderson, and Jianzhong Wu.
\newblock Solvent effect on the pore-size dependence of an organic electrolyte
  supercapacitor.
\newblock \emph{The journal of physical chemistry letters}, 3\penalty0
  (13):\penalty0 1727--1731, 2012.

\bibitem[Jiang et~al.(2014)Jiang, Cao, Jiang, and Wu]{jiang2014time}
Jian Jiang, Dapeng Cao, De-en Jiang, and Jianzhong Wu.
\newblock Time-dependent density functional theory for ion diffusion in
  electrochemical systems.
\newblock \emph{Journal of Physics: Condensed Matter}, 26\penalty0
  (28):\penalty0 284102, 2014.

\bibitem[Keiser et~al.(1976)Keiser, Beccu, and Gutjahr]{keiser1976abschatzung}
H~Keiser, K~D Beccu, and M~A Gutjahr.
\newblock Absch{\"a}tzung der porenstruktur por{\"o}ser elektroden aus
  impedanzmessungen.
\newblock \emph{Electrochimica Acta}, 21\penalty0 (8):\penalty0 539--543, 1976.

\bibitem[Kierlik and Rosinberg(1991)]{kierlik1991density}
E~Kierlik and ML~Rosinberg.
\newblock Density-functional theory for inhomogeneous fluids: adsorption of
  binary mixtures.
\newblock \emph{Physical Review A}, 44\penalty0 (8):\penalty0 5025, 1991.

\bibitem[Kilic et~al.(2007{\natexlab{a}})Kilic, Bazant, and
  Ajdari]{kilic2007steric1}
Mustafa~Sabri Kilic, Martin~Z Bazant, and Armand Ajdari.
\newblock Steric effects in the dynamics of electrolytes at large applied
  voltages. {I}. {D}ouble-layer charging.
\newblock \emph{Physical review E}, 75\penalty0 (2):\penalty0 021502,
  2007{\natexlab{a}}.

\bibitem[Kilic et~al.(2007{\natexlab{b}})Kilic, Bazant, and
  Ajdari]{kilic2007steric2}
Mustafa~Sabri Kilic, Martin~Z Bazant, and Armand Ajdari.
\newblock Steric effects in the dynamics of electrolytes at large applied
  voltages. {II}. {M}odified {P}oisson--{N}ernst--{P}lanck equations.
\newblock \emph{Physical review E}, 75\penalty0 (2):\penalty0 021503,
  2007{\natexlab{b}}.

\bibitem[Kirk et~al.(2020)Kirk, Evans, Please, and Chapman]{kirk2020modelling}
Toby~L Kirk, Jack Evans, Colin~P Please, and S~Jonathan Chapman.
\newblock Modelling electrode heterogeneity in lithium-ion batteries: unimodal
  and bimodal particle-size distributions.
\newblock \emph{arXiv preprint arXiv:2006.12208}, 2020.

\bibitem[Knepley et~al.(2010)Knepley, Karpeev, Davidovits, Eisenberg, and
  Gillespie]{knepley2010efficient}
Matthew~G Knepley, Dmitry~A Karpeev, Seth Davidovits, Robert~S Eisenberg, and
  Dirk Gillespie.
\newblock An efficient algorithm for classical density functional theory in
  three dimensions: {I}onic solutions.
\newblock \emph{The Journal of chemical physics}, 132\penalty0 (12):\penalty0
  124101, 2010.

\bibitem[Kondrat and Kornyshev(2013)]{kondrat2013charging}
S~Kondrat and A~Kornyshev.
\newblock Charging dynamics and optimization of nanoporous supercapacitors.
\newblock \emph{The Journal of Physical Chemistry C}, 117\penalty0
  (24):\penalty0 12399--12406, 2013.

\bibitem[Kondrat et~al.(2014)Kondrat, Wu, Qiao, and
  Kornyshev]{kondrat2014accelerating}
Svyatoslav Kondrat, Peng Wu, Rui Qiao, and Alexei~A Kornyshev.
\newblock Accelerating charging dynamics in subnanometre pores.
\newblock \emph{Nature materials}, 13\penalty0 (4):\penalty0 387--393, 2014.

\bibitem[Kornyshev(2007)]{doi:10.1021/jp067857o}
Alexei~A. Kornyshev.
\newblock Double-{L}ayer in {I}onic {L}iquids: a {P}aradigm {C}hange?
\newblock \emph{The Journal of Physical Chemistry B}, 111\penalty0
  (20):\penalty0 5545--5557, 2007.
\newblock \doi{10.1021/jp067857o}.

\bibitem[Kroupa et~al.(2016)Kroupa, Offer, and Kosek]{kroupa2016modelling}
Martin Kroupa, Gregory~J Offer, and Juraj Kosek.
\newblock Modelling of supercapacitors: {F}actors influencing performance.
\newblock \emph{Journal of The Electrochemical Society}, 163\penalty0
  (10):\penalty0 A2475, 2016.

\bibitem[Li and Huang(2021)]{li2021impedance}
Chen~Kun Li and Jun Huang.
\newblock Impedance response of electrochemical interfaces: part {I}. {E}xact
  analytical expressions for ideally polarizable electrodes.
\newblock \emph{Journal of the Electrochemical Society}, 167\penalty0
  (16):\penalty0 166517, 2021.

\bibitem[Lian et~al.(2016{\natexlab{a}})Lian, Liu, Van~Aken, Gogotsi,
  Wesolowski, Liu, Jiang, and Wu]{lian2016enhancing}
C~Lian, K~Liu, Katherine~L Van~Aken, Y~Gogotsi, David~J Wesolowski, HL~Liu,
  DE~Jiang, and JZ~Wu.
\newblock Enhancing the capacitive performance of electric double-layer
  capacitors with ionic liquid mixtures.
\newblock \emph{ACS Energy Letters}, 1\penalty0 (1):\penalty0 21--26,
  2016{\natexlab{a}}.

\bibitem[Lian et~al.(2016{\natexlab{b}})Lian, Zhao, Liu, and Wu]{lian2016time}
Cheng Lian, Shuangliang Zhao, Honglai Liu, and Jianzhong Wu.
\newblock Time-dependent density functional theory for the charging kinetics of
  electric double layer containing room-temperature ionic liquids.
\newblock \emph{The Journal of chemical physics}, 145\penalty0 (20):\penalty0
  204707, 2016{\natexlab{b}}.

\bibitem[Lim et~al.(2007)Lim, Whitcomb, Boyd, and Varghese]{LIM2007159}
Jongil Lim, John Whitcomb, James Boyd, and Julian Varghese.
\newblock Transient finite element analysis of electric double layer using
  {N}ernst--{P}lanck--{P}oisson equations with a modified {S}tern layer.
\newblock \emph{Journal of Colloid and Interface Science}, 305\penalty0
  (1):\penalty0 159 -- 174, 2007.
\newblock ISSN 0021-9797.
\newblock \doi{https://doi.org/10.1016/j.jcis.2006.08.049}.
\newblock URL
  \url{http://www.sciencedirect.com/science/article/pii/S0021979706007788}.

\bibitem[Liu et~al.(2017)Liu, Lian, Henderson, and Wu]{liu2017impurity}
Kun Liu, Cheng Lian, Douglas Henderson, and Jianzhong Wu.
\newblock Impurity effects on ionic-liquid-based supercapacitors.
\newblock \emph{Molecular Physics}, 115\penalty0 (4):\penalty0 454--464, 2017.

\bibitem[Lutsko(2010)]{lutsko2010recent}
James~F Lutsko.
\newblock Recent developments in classical density functional theory.
\newblock \emph{Advances in chemical physics}, 144:\penalty0 1, 2010.

\bibitem[Lutsko and Lam(2018)]{lutsko2018classical}
James~F Lutsko and Julien Lam.
\newblock Classical density functional theory, unconstrained crystallization,
  and polymorphic behavior.
\newblock \emph{Physical Review E}, 98\penalty0 (1):\penalty0 012604, 2018.

\bibitem[Marconi and Tarazona(1999)]{marconi1999dynamic}
Umberto Marini~Bettolo Marconi and Pedro Tarazona.
\newblock Dynamic density functional theory of fluids.
\newblock \emph{The Journal of chemical physics}, 110\penalty0 (16):\penalty0
  8032--8044, 1999.

\bibitem[Medasani et~al.(2014)Medasani, Ovanesyan, Thomas, Sushko, and
  Marucho]{medasani2014ionic}
Bharat Medasani, Zaven Ovanesyan, Dennis~G Thomas, Maria~L Sushko, and Marcelo
  Marucho.
\newblock Ionic asymmetry and solvent excluded volume effects on spherical
  electric double layers: A density functional approach.
\newblock \emph{The Journal of chemical physics}, 140\penalty0 (20):\penalty0
  204510, 2014.

\bibitem[Narski and Ottaviani(2014)]{narski2014asymptotic}
Jacek Narski and Maurizio Ottaviani.
\newblock Asymptotic preserving scheme for strongly anisotropic parabolic
  equations for arbitrary anisotropy direction.
\newblock \emph{Computer Physics Communications}, 185\penalty0 (12):\penalty0
  3189--3203, 2014.

\bibitem[Neal et~al.(2017)Neal, Wesolowski, Henderson, and Wu]{neal2017ion}
Justin~N Neal, David~J Wesolowski, Douglas Henderson, and Jianzhong Wu.
\newblock Ion distribution and selectivity of ionic liquids in microporous
  electrodes.
\newblock \emph{The Journal of chemical physics}, 146\penalty0 (17):\penalty0
  174701, 2017.

\bibitem[Noda et~al.(2001)Noda, Hayamizu, and Watanabe]{noda2001pulsed}
Akihiro Noda, Kikuko Hayamizu, and Masayoshi Watanabe.
\newblock Pulsed-gradient spin- echo 1h and 19f nmr ionic diffusion
  coefficient, viscosity, and ionic conductivity of non-chloroaluminate
  room-temperature ionic liquids.
\newblock \emph{The Journal of Physical Chemistry B}, 105\penalty0
  (20):\penalty0 4603--4610, 2001.

\bibitem[Qing et~al.(2019)Qing, Li, Tang, Zhang, Han, and
  Zhao]{qing2019dynamic}
Leying Qing, Yu~Li, Weiqiang Tang, Duo Zhang, Yongsheng Han, and Shuangliang
  Zhao.
\newblock Dynamic adsorption of ions into like-charged nanospace: A dynamic
  density functional theory study.
\newblock \emph{Langmuir}, 35\penalty0 (12):\penalty0 4254--4262, 2019.

\bibitem[Qing et~al.(2020)Qing, Lei, Zhao, Qiu, Ma, Xu, and
  Zhao]{qing2020effects}
Leying Qing, Jun Lei, Teng Zhao, Genlong Qiu, Manman Ma, Zhenli Xu, and
  Shuangliang Zhao.
\newblock Effects of kinetic dielectric decrement on ion diffusion and
  capacitance in electrochemical systems.
\newblock \emph{Langmuir}, 2020.

\bibitem[Qing et~al.(2021)Qing, Zhao, and Wang]{qingsurface}
Leying Qing, Shuangliang Zhao, and Zhen-Gang Wang.
\newblock Surface charge density in electrical double layer capacitors with
  nanoscale cathode--anode separation.
\newblock \emph{The Journal of Physical Chemistry B}, 125\penalty0
  (2):\penalty0 625--636, 2021.
\newblock \doi{10.1021/acs.jpcb.0c09332}.

\bibitem[Roling and Dr{\"u}schler(2012)]{ROLING2012526}
Bernhard Roling and Marcel Dr{\"u}schler.
\newblock Comments on ``{I}ntrinsic limitations of impedance measurements in
  determining electric double layer capacitances'' by {H. Wang and L. Pilon
  [Electrochim. Acta 63 (2012) 55}].
\newblock \emph{Electrochimica Acta}, 76:\penalty0 526 -- 528, 2012.
\newblock ISSN 0013-4686.
\newblock \doi{https://doi.org/10.1016/j.electacta.2012.03.180}.
\newblock URL
  \url{http://www.sciencedirect.com/science/article/pii/S0013468612006329}.

\bibitem[Rosenfeld(1989)]{rosenfeld1989free}
Yaakov Rosenfeld.
\newblock Free-energy model for the inhomogeneous hard-sphere fluid mixture and
  density-functional theory of freezing.
\newblock \emph{Physical review letters}, 63\penalty0 (9):\penalty0 980, 1989.

\bibitem[Rosenfeld(1993)]{rosenfeld1993free}
Yaakov Rosenfeld.
\newblock Free energy model for inhomogeneous fluid mixtures: Yukawa-charged
  hard spheres, general interactions, and plasmas.
\newblock \emph{The Journal of chemical physics}, 98\penalty0 (10):\penalty0
  8126--8148, 1993.

\bibitem[Roth et~al.(2002)Roth, Evans, Lang, and Kahl]{roth2002fundamental}
R~Roth, R~Evans, A~Lang, and G~Kahl.
\newblock Fundamental measure theory for hard-sphere mixtures revisited: the
  {W}hite {B}ear version.
\newblock \emph{Journal of Physics: Condensed Matter}, 14\penalty0
  (46):\penalty0 12063, 2002.

\bibitem[Roth(2010)]{roth2010fundamental}
Roland Roth.
\newblock Fundamental measure theory for hard-sphere mixtures: a review.
\newblock \emph{Journal of Physics: Condensed Matter}, 22\penalty0
  (6):\penalty0 063102, 2010.

\bibitem[Roth and Gillespie(2016)]{roth2016shells}
Roland Roth and Dirk Gillespie.
\newblock Shells of charge: a density functional theory for charged hard
  spheres.
\newblock \emph{Journal of Physics: Condensed Matter}, 28\penalty0
  (24):\penalty0 244006, 2016.

\bibitem[Russo et~al.(2020)Russo, Perez, Dur{\'a}n-Olivencia, Yatsyshin,
  Carrillo, and Kalliadasis]{russo2020finite}
Antonio Russo, Sergio~P Perez, Miguel~A Dur{\'a}n-Olivencia, Peter Yatsyshin,
  Jos{\'e}~A Carrillo, and Serafim Kalliadasis.
\newblock A finite-volume method for fluctuating dynamical density functional
  theory.
\newblock \emph{Journal of Computational Physics}, page 109796, 2020.

\bibitem[Schmidt and Brader(2013)]{schmidt2013power}
Matthias Schmidt and Joseph~M Brader.
\newblock Power functional theory for {B}rownian dynamics.
\newblock \emph{The Journal of chemical physics}, 138\penalty0 (21):\penalty0
  214101, 2013.

\bibitem[Song and Bazant(2012)]{song2012effects}
Juhyun Song and Martin~Z Bazant.
\newblock Effects of nanoparticle geometry and size distribution on diffusion
  impedance of battery electrodes.
\newblock \emph{Journal of The Electrochemical Society}, 160\penalty0
  (1):\penalty0 A15, 2012.

\bibitem[Stopper et~al.(2018)Stopper, Thorneywork, Dullens, and
  Roth]{stopper2018bulk}
Daniel Stopper, Alice~L Thorneywork, Roel~PA Dullens, and Roland Roth.
\newblock Bulk dynamics of {B}rownian hard disks: {D}ynamical density
  functional theory versus experiments on two-dimensional colloidal hard
  spheres.
\newblock \emph{The Journal of chemical physics}, 148\penalty0 (10):\penalty0
  104501, 2018.

\bibitem[Tang and Wu(2003)]{tang2003density}
Yiping Tang and Jianzhong Wu.
\newblock A density-functional theory for bulk and inhomogeneous
  {L}ennard--{J}ones fluids from the energy route.
\newblock \emph{The Journal of chemical physics}, 119\penalty0 (14):\penalty0
  7388--7397, 2003.

\bibitem[Tarazona et~al.(2008)Tarazona, Cuesta, and
  Mart{\'\i}nez-Rat{\'o}n]{tarazona2008density}
Pedro Tarazona, Jos{\'e}~A Cuesta, and Yuri Mart{\'\i}nez-Rat{\'o}n.
\newblock Density functional theories of hard particle systems.
\newblock In \emph{Theory and Simulation of Hard-Sphere Fluids and Related
  Systems}, pages 247--341. Springer, 2008.

\bibitem[te~Vrugt et~al.(2020)te~Vrugt, L{\"o}wen, and
  Wittkowski]{te2020classical}
Michael te~Vrugt, Hartmut L{\"o}wen, and Raphael Wittkowski.
\newblock Classical dynamical density functional theory: from fundamentals to
  applications.
\newblock \emph{Advances in Physics}, 69\penalty0 (2):\penalty0 121--247, 2020.

\bibitem[Terrones et~al.(2016)Terrones, Kiley, and
  Elliott]{terrones2016enhanced}
Jeronimo Terrones, Patrick~J Kiley, and James~A Elliott.
\newblock Enhanced ordering reduces electric susceptibility of liquids confined
  to graphene slit pores.
\newblock \emph{Scientific reports}, 6\penalty0 (1):\penalty0 1--11, 2016.

\bibitem[Valisk{\'o} et~al.(2018)Valisk{\'o}, Krist{\'o}f, Gillespie, and
  Boda]{valisko2018systematic}
M{\'o}nika Valisk{\'o}, Tam{\'a}s Krist{\'o}f, Dirk Gillespie, and Dezs{\H{o}}
  Boda.
\newblock A systematic monte carlo simulation study of the primitive model
  planar electrical double layer over an extended range of concentrations,
  electrode charges, cation diameters and valences.
\newblock \emph{AIP Advances}, 8\penalty0 (2):\penalty0 025320, 2018.

\bibitem[Verbrugge and Liu(2005)]{verbrugge2005microstructural}
Mark~W Verbrugge and Ping Liu.
\newblock Microstructural analysis and mathematical modeling of electric
  double-layer supercapacitors.
\newblock \emph{Journal of the Electrochemical Society}, 152\penalty0
  (5):\penalty0 D79, 2005.

\bibitem[Voukadinova and Gillespie(2019)]{voukadinova2019energetics}
Adelina Voukadinova and Dirk Gillespie.
\newblock Energetics of counterion adsorption in the electrical double layer.
\newblock \emph{The Journal of chemical physics}, 150\penalty0 (15):\penalty0
  154706, 2019.

\bibitem[Voukadinova et~al.(2018)Voukadinova, Valisk{\'o}, and
  Gillespie]{voukadinova2018assessing}
Adelina Voukadinova, M{\'o}nika Valisk{\'o}, and Dirk Gillespie.
\newblock Assessing the accuracy of three classical density functional theories
  of the electrical double layer.
\newblock \emph{Physical Review E}, 98\penalty0 (1):\penalty0 012116, 2018.

\bibitem[Wang and Pilon(2012{\natexlab{a}})]{WANG2012529}
Hainan Wang and Laurent Pilon.
\newblock Reply to comments on ``{Intrinsic limitations of impedance
  measurements in determining electric double layer capacitances'' by H. Wang,
  L. Pilon [Electrochimica Acta 63 (2012) 55]}.
\newblock \emph{Electrochimica Acta}, 76:\penalty0 529 -- 531,
  2012{\natexlab{a}}.
\newblock ISSN 0013-4686.
\newblock \doi{https://doi.org/10.1016/j.electacta.2012.05.039}.
\newblock URL
  \url{http://www.sciencedirect.com/science/article/pii/S0013468612008043}.

\bibitem[Wang and Pilon(2012{\natexlab{b}})]{wang2012intrinsic}
Hainan Wang and Laurent Pilon.
\newblock Intrinsic limitations of impedance measurements in determining
  electric double layer capacitances.
\newblock \emph{Electrochimica Acta}, 63:\penalty0 55--63, 2012{\natexlab{b}}.

\bibitem[Wang et~al.(2011)Wang, Liu, and Neretnieks]{wang2011weighted}
Zhao Wang, Longcheng Liu, and Ivars Neretnieks.
\newblock The weighted correlation approach for density functional theory: a
  study on the structure of the electric double layer.
\newblock \emph{Journal of Physics: Condensed Matter}, 23\penalty0
  (17):\penalty0 175002, 2011.

\bibitem[Wu et~al.(2011)Wu, Jiang, Jiang, Jin, and Henderson]{wu2011classical}
Jianzhong Wu, Tao Jiang, De-en Jiang, Zhehui Jin, and Douglas Henderson.
\newblock A classical density functional theory for interfacial layering of
  ionic liquids.
\newblock \emph{Soft Matter}, 7\penalty0 (23):\penalty0 11222--11231, 2011.

\bibitem[Yu et~al.(2017)Yu, Jabeen, Eckmann, Ayyaswamy, and
  Radhakrishnan]{yu2017microstructure}
Hsiu-Yu Yu, Zahera Jabeen, David~M Eckmann, Portonovo~S Ayyaswamy, and Ravi
  Radhakrishnan.
\newblock Microstructure of flow-driven suspension of hardspheres in
  cylindrical confinement: a dynamical density functional theory and {M}onte
  {C}arlo study.
\newblock \emph{Langmuir}, 33\penalty0 (42):\penalty0 11332--11344, 2017.

\bibitem[Yu and Wu(2002)]{yu2002structures}
Yang-Xin Yu and Jianzhong Wu.
\newblock Structures of hard-sphere fluids from a modified fundamental-measure
  theory.
\newblock \emph{The Journal of chemical physics}, 117\penalty0 (22):\penalty0
  10156--10164, 2002.

\end{thebibliography}

\end{document}